\documentclass{emulateapj}
\usepackage{amssymb,amsfonts,amsmath,amstext,amsgen,amsopn,amsxtra,indentfirst,lscape}
\shorttitle{MHD Shallow Water Waves}
\shortauthors{Heng \& Spitkovsky}
\begin{document}

\title{Magnetohydrodynamic Shallow Water Waves: Linear Analysis}

\author{Kevin Heng\altaffilmark{1} \& Anatoly Spitkovsky\altaffilmark{2}}

\altaffiltext{1}{Institute for Advanced Study, School of Natural Sciences, Einstein Drive, Princeton, NJ 08540; heng@ias.edu}

\altaffiltext{2}{Department of Astrophysical Sciences, Peyton Hall, Princeton University, Princeton, NJ 08544; anatoly@astro.princeton.edu}

\begin{abstract}
We present a linear analysis of inviscid, incompressible, magnetohydrodynamic (MHD) shallow water systems.  In spherical geometry, a generic property of such systems is the existence of five wave modes.  Three of them (two magneto-Poincar\'{e} modes and one magneto-Rossby mode) are previously known.  The other two wave modes are strongly influenced by the magnetic field and rotation, and have substantially lower angular frequencies; as such, we term them ``magnetostrophic modes''.  We obtain analytical functions for the velocity, height and magnetic field perturbations in the limit that the magnitude of the MHD analogue of Lamb's parameter is large.  On a sphere, the magnetostrophic modes reside near the poles, while the other modes are equatorially confined. Magnetostrophic modes may be an ingredient in explaining the frequency drifts observed in Type I X-ray bursts from neutron stars.
\end{abstract}


\section{Introduction}
\label{sect:intro}

Shallow water wave systems are those in which the transverse length scales considered are much larger than the water height.  They have been extensively studied, have a wide range of physical applications, and have been used to understand terrestrial and planetary systems (Matsuno [1966]; Gill [1982], hereafter G82; Pedlosky [1987]; Braginsky [1998]; Holton [2004]; and Kundu \& Cohen [2004], hereafter KC04).  Shallow water equations capture the large scale dynamics of thinly-stratified atmospheres under the influence of rotation and thermal forcing. 

In astrophysical settings, shallow water models were used to study the spread of accreted matter onto neutron stars (Inogamov \& Sunyaev 1999), and the ignition and propagation of Type I X-ray bursts in neutron star atmospheres (Spitkovsky, Levin \& Ushomirsky 2002). In addition to fast rotation and strong gravity, many neutron stars in accreting systems also possess appreciable ($\lesssim 10^9$G) magnetic fields. Tension in the fields induced by horizontal motions in the atmosphere can affect the dynamics, and thus it is interesting to study the effect of frozen-in magnetic fields on the behavior of the atmosphere. 

Gilman (2000) pioneered the use of magnetohydrodynamic (MHD) shallow water systems for studying the solar tachocline, which inspired several follow-up studies (e.g., Schecter, Boyd \& Gilman 2001; Zaqarashvili et al. 2007, 2009). Most works on MHD shallow water systems consider initially toroidal magnetic fields and/or slow rotators, as is applicable for tachocline research. In neutron star-related applications, however, it makes sense to consider initially vertical (or radial) magnetic fields in the presence of fast rotation. We study this case in the present paper. 

The fundamental governing equation for shallow water waves on a sphere is known as Laplace's tidal equation.  By linearizing this equation, one can obtain solutions to the water height and velocity perturbations, as well as dispersion relations for the angular frequencies of the waves.  In rotating systems, it was realized (Longuet-Higgins 1965) that an important quantity in such studies is Lamb's parameter,
\begin{equation}
\epsilon \equiv \frac{1}{{\cal R}^2},
\end{equation}
where ${\cal R}$ is the Rossby number\footnote{Defined as the ratio of inertia to Coriolis forces on the scale of the planet or star.}.  Physically, shallow water systems with large values of $\epsilon$ are ``fast rotators''.  In a seminal paper, Longuet-Higgins (1968, hereafter LH68) explored solutions to Laplace's tidal equation for a wide range of values for $\epsilon$ and demonstrated that analytical forms exist for $\epsilon \rightarrow 0$ and $\vert \epsilon \vert \rightarrow \infty$.  While $\epsilon < 0$ solutions may appear unphysical, LH68 realized that they are relevant to the study of forced oscillations.

\begin{figure}
\begin{center}
\includegraphics[width=3.2in]{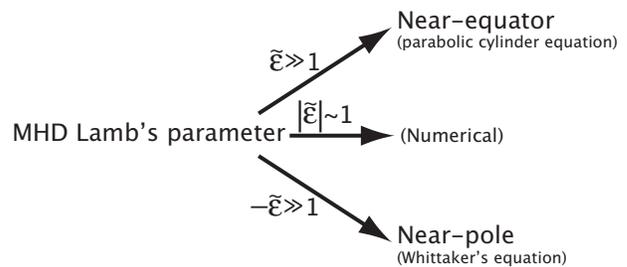}
\end{center}
\caption{Qualitative behaviour of the wave modes, as controlled by the MHD Lamb's parameter.  The governing equation for the latitudinal velocity perturbation, in the respective limits, is given in parentheses.}
\label{fig:lamb}
\vspace{0.1in}
\end{figure}

When magnetic fields are considered, one needs to instead examine the MHD Lamb's parameter,
\begin{equation}
\tilde{\epsilon} = {\cal I}\left(B_z, \Omega \right) \epsilon.
\end{equation}
The function ${\cal I}={\cal I}(B_z, \Omega)$, which we will define later, depends on the rotation frequency of the system $\Omega$ and the magnetic field strength $B_z$; for hydrodynamic waves, ${\cal I}=1$.  Unlike in the case of hydrodynamic systems, $\tilde{\epsilon}<0$ waves are directly relevant in MHD --- for $-\tilde{\epsilon} \gg 1$, they reside near the poles on a spherical surface.  By contrast, $\tilde{\epsilon} \gg 1$ waves reside near the equator.  Figure \ref{fig:lamb} shows a schematic demarcating the asymptotic behaviour of the wave solutions and the limiting equation that governs the latitudinal velocity perturbation.  

In this paper, our goal is to demonstrate that the methods of LH68 can be generalized to obtain analytical, asymptotic ($\vert \tilde{\epsilon} \vert \gg 1$) solutions to the MHD analogue of Laplace's tidal equation.  These solutions can then be used as a guide towards obtaining $\vert \tilde{\epsilon} \vert \sim 1$ solutions, which must be computed numerically.  Throughout the study, we shall adopt a radial magnetic field for simplicity (and algebraic amenability).  Our results can be straightforwardly generalized for arbitrary magnetic field configurations.

Readers unfamiliar with the classical shallow water analysis are referred to Appendices \ref{sect:classical_cartesian} and \ref{sect:classical_spherical}, where the hydrodynamic treatment is presented in Cartesian and spherical geometry, respectively.  In \S\ref{sect:mhd_cartesian}, we add magnetic fields to the Cartesian analysis.  Our efforts culminate in \S\ref{sect:mhd_spherical}, where we explore MHD solutions on a sphere.  We discuss the implications of our results in \S\ref{sect:discussion}.  A concise summary of the paper is presented in \S\ref{sect:summary}.

\section{MHD Shallow Water Waves: Cartesian Coordinates}
\label{sect:mhd_cartesian}

Consider a Cartesian coordinate system in which a shallow layer of water of height $h = h(x,y,t)$ resides at $z \ge 0$ on an infinite plane.  The fluid moves with a velocity $\vec{v} = (v_x,v_y,v_z)$ such that
\begin{equation}
\begin{split}
&v_x = v_x(x,y,t),\\
&v_y = v_y(x,y,t).\\
\end{split}
\end{equation}
The fluid is assumed to be inviscid and incompressible.

\subsection{$f$-Plane Treatment (Magneto-Poincar\'{e} and Geostrophic Waves)}
\label{subsect:fplane_mhd}

\subsubsection{Equations}

Consider the Euler equation with the Coriolis force and magnetic tension terms (e.g., Draine 1986):
\begin{equation}
\frac{\partial \vec{v}}{\partial t} + \vec{v}.\nabla \vec{v} = -\frac{1}{\rho} \nabla P - 2\left( \vec{\Omega} \times \vec{v} \right) + \frac{1}{4 \pi \rho} \vec{B}.\nabla\vec{B}.
\label{eq:mhd_laplace}
\end{equation}
Incompressibility implies (see Appendix \ref{subsect:gravity})
\begin{equation}
\frac{\partial h}{\partial t} + \nabla.\left(h \vec{v}\right) = 0.
\end{equation}
The induction equation, in the ideal MHD limit, provides an additional equation of motion:
\begin{equation}
\frac{\partial \vec{B}}{\partial t} = \nabla \times \left( \vec{v} \times \vec{B} \right).
\label{eq:induction}
\end{equation}

We consider small perturbations to the water height, as well as to the $x$- and $y$-components of the velocity and magnetic field:
\begin{equation}
\begin{split}
&v_x = V_x + v^\prime_x,\\
&v_y = V_y + v^\prime_y,\\
&h = H + h^\prime,\\
&B_x = B_{x_0} + b_x,\\
&B_y = B_{y_0} + b_y.\\
\end{split}
\end{equation}
The field is initially vertical (we will set $B_{x_0} = B_{y_0} = 0$ later), and is frozen at the bottom of the atmosphere. The largest gradient of the field is due to the vertical shear in the layer, which we approximate as:
\begin{equation}
\frac{\partial B_j}{\partial z} = - \frac{B_j}{H},
\label{eq:b_gradient_approx}
\end{equation}
where the index $j$ represents the set $j = \{x,y\}$; $B_z$ is taken to be constant.  The net effect of these approximations is to introduce restoring tension forces that pull the fluid elements back into their original horizontal position.  In the shallow water formalism, one deals with height-averaged horizontal velocities and accelerations.  Collectively, it is then reasonable to take the restoring force (per unit volume) to have a magnitude $B^2_z/4\pi H$ and a direction opposite to that of the horizontal displacement. 

The following wave solutions are sought:
\begin{equation}
\begin{split}
&v^\prime_x = v_{x_0} \exp{i\left( k_x x + k_y y - \omega t \right)},\\
&v^\prime_y = v_{y_0} \exp{i\left( k_x x + k_y y - \omega t \right)},\\
&h^\prime = h_0 \exp{i\left( k_x x + k_y y - \omega t \right)},\\
&b_x = b_{x_0} \exp{i\left( k_x x + k_y y - \omega t \right)},\\
&b_y = b_{y_0} \exp{i\left( k_x x + k_y y - \omega t \right)},\\
\end{split}
\end{equation}
where the wave vector is $\vec{k} = (k_x,k_y,0)$.  Linearization of the equations of motion yields a $5 \times 5$ matrix (see Appendix \ref{subsect:gravity}):
\[ \hat{A} = \left( \begin{array}{ccccc}
k_x H & k_y H & A_0 & 0 & 0\\
iA_0 & -2\Omega & igk_x & A_{2-} & 0\\
2\Omega & iA_0 & ig k_y & 0 & A_{2-}\\
A_3 & iB_{x_0} k_y & 0 & i\left( \omega - k_y V_y \right) & ik_y V_x\\
iB_{y_0} k_x & A_4 & 0 & ik_x V_y & i\left( \omega - k_x V_x \right) \end{array} \right),\]
where
\begin{equation}
\begin{split}
&A_0 \equiv -\omega + k_x V_x + k_y V_y,\\
&A_1 \equiv \frac{B_z}{4 \pi \rho H},\\
&A_{2\pm} \equiv A_1 \pm \frac{i}{4 \pi \rho}\left( B_{x_0} k_x + B_{y_0} k_y \right),\\
&A_3 \equiv 4 \pi \rho A_{2+} + iB_{x_0} k_x,\\
&A_4 \equiv 4 \pi \rho A_{2+} + iB_{y_0} k_y.\\
\end{split}
\end{equation}
A practical note about evaluating det$\hat{A}$ is that one is free to permute the rows of $\hat{A}$, and they should be arranged in a way so as to minimize the number of sub-determinant (i.e., of $4 \times 4$ matrices) evaluations.

\subsubsection{Dispersion Relation}
\label{subsect:dispersion_fplane_mhd}

Evaluating det$\hat{A}=0$, setting $V_x = V_y = B_{x_0} = B_{y_0} = 0$ and 
collecting terms yields the dispersion relation:
\begin{equation}
\begin{split}
&\omega^4 - \omega^2 \left[ gHk^2 + 4\Omega^2 + 2(v_A/H)^2 \right]\\
&+ (v_A/H)^2\left[gHk^2 + (v_A/H)^2 \right] = 0,\\
\end{split}
\label{eq:dispersion_fplane_mhd}
\end{equation}
where $v_A \equiv B_z/2\sqrt{\pi \rho}$ is the Alfv\'{e}n speed.  Its solution is
\begin{equation}
\begin{split}
\omega^2 =& \frac{gHk^2}{2} + 2\Omega^2 + \frac{B^2_z}{4\pi \rho H^2}\\
&\pm \frac{1}{2}\sqrt{gHk^2\left( gHk^2 + 8\Omega^2 \right) + 16\Omega^2\left( \Omega^2 + \frac{B^2_z}{4\pi \rho H^2} \right)}.\\
\end{split}
\label{eq:dispersion_fplane_mhd_solution}
\end{equation}

Some intuition can be developed by examining the solutions in $\Omega=0$ limit:
\begin{equation}
\omega^2  =
\begin{cases}
gHk^2 + (v_A/H)^2,\\
(v_A/H)^2.\\
\end{cases}
\end{equation}
The first mode is the longitudinal ``magnetogravity wave'' (Schecter, Boyd \& Gilman 2001), where the restoring force is a combination of the rising fluid height and magnetic pressure increase due to the compressive motion of the gravity wave. The second mode, which is non-dispersive, represents the Alfv\'{e}n wave with no height perturbation and torsional oscillation in the transverse direction, whose restoring force is magnetic tension. 

When we revert to rotating systems, both modes begin to couple longitudinal and transverse velocities through the Coriolis force, but have different polarizations. The higher-frequency mode in equation (\ref{eq:dispersion_fplane_mhd_solution}) is the rotationally-modified magnetogravity wave, or ``magneto-Poincar\'{e}'' wave --- the magnetic tension term in equation (\ref{eq:mhd_laplace}) adds in phase with the Coriolis force, thus enhancing the restoring force and speeding up the oscillation. In the lower-frequency mode from equation (\ref{eq:dispersion_fplane_mhd_solution}), magnetic tension tries to balance the Coriolis force. To emphasize this balance, we call this the ``magnetostrophic mode''. 

Both magneto-Poincar\'{e} and magnetostrophic modes come in eastward- and westward-propagating varieties, giving the four modes as required by equation (\ref{eq:dispersion_fplane_mhd}). The eastward- and westward-propagating waves of each branch have the same frequency by absolute value. As the magnetic field is turned off, the magnetostrophic mode disappears, while the Poincar\'{e} mode (or rotationally-modified gravity wave) survives as expected. 

\subsubsection{Length Scales}
\label{subsect:magnetostrophic}

In hydrodynamic systems, rotational effects become important at wavelengths of $\lambda \equiv 1/k \gtrsim \lambda_{\rm R}$, where $k \equiv \vert \vec{k} \vert = \sqrt{k^2_x + k^2_y}$ and
\begin{equation}
\lambda_{\rm R} \equiv \frac{\sqrt{gH}}{2\Omega}
\label{eq:rossby_radius}
\end{equation}
is the ``Rossby radius of deformation'' (e.g., chapter 7 of G82).  The Rossby radius can also be obtained by arguing that there exists a radius at which the radial fluid flow is diverted by the Coriolis force, i.e., $\lambda_{\rm R} \Omega \sim c_0$, where
\begin{equation}
c_0 \equiv \sqrt{gH}
\end{equation}
is the shallow water wave speed.  Physically, an adjustment to ``geostrophic balance'' occurs at the Rossby radius on a timescale $\sim 1/\Omega$ (pg. 201 of G82).

In MHD systems, even without explicitly calculating the group velocity, one can see from equation (\ref{eq:dispersion_fplane_mhd}) that if $v_A \sim \Omega H$, we have
\begin{equation}
\lambda_{\rm RB} \equiv \sqrt{ \frac{gH}{4\Omega^2 + 2(v_A/H)^2} }.
\end{equation}
The quantity $\lambda_{\rm RB}$ can be understood in the following manner: if the forces due to rotation and magnetic tension are equally important, then at wavelengths of $\lambda \sim \lambda_{\rm RB}$ they balance out the effect of gravity.

A somewhat more relevant quantity to define is the Alfv\'{e}n radius, $\lambda_{\rm B}$.  In the absence of rotation, the forces due to magnetic tension and gravity balance at $\lambda \sim \lambda_{\rm B}$, where
\begin{equation}
\lambda_{\rm B} \equiv \lambda_{\rm RB}\left(\Omega=0\right) = \sqrt{gH} ~\left( \frac{B_z}{\sqrt{2 \pi \rho}} \right)^{-1} ~H.
\label{eq:alfven_radius}
\end{equation}
The preceding expression can be approximately obtained by arguing that $\lambda_{\rm B} \sim c_0 H/v_A$.
The relative importance of rotation and magnetic tension can be judged from the ratio $\lambda_{\rm R}/\lambda_{\rm B}$. 
\subsection{$\beta$-Plane Treatment (Magneto-Poincar\'{e} and Magneto-Rossby Waves)}
\label{subsect:betaplane_mhd}

By analogy with hydrodynamic systems, we term the very slow waves with large wavelength ``magneto-Rossby waves''.  Generalizing the analysis in Appendix \ref{subsect:betaplane}, we set $V_x = V_y = B_{x_0} = B_{y_0} = 0$ and consider the following set of linearized equations:
\begin{equation}
\begin{split}
&\frac{\partial v^\prime_x}{\partial t} + g \frac{\partial h^\prime}{\partial x} - f v^\prime_y + A_1 b_x =0,\\
&\frac{\partial v^\prime_y}{\partial t} + g \frac{\partial h^\prime}{\partial y} + f v^\prime_x + A_1 b_y = 0,\\
&\frac{\partial h^\prime}{\partial t} + \left( \frac{\partial v^\prime_x}{\partial x} + \frac{\partial v^\prime_y}{\partial y} \right)H = 0,\\
&\frac{\partial b_x}{\partial t} = \frac{v^\prime_x B_z}{H},\\
&\frac{\partial b_y}{\partial t} = \frac{v^\prime_y B_z}{H},\\
\end{split}
\label{eq:linear_betaplane_mhd}
\end{equation}
where $f = 2\Omega \sin{\Theta} + \beta y$ is the Coriolis parameter and $\Theta$ denotes the latitude.  We differentiate the first equation in (\ref{eq:linear_betaplane_mhd}) with respect to $y$ and seek wave solutions from the entire set of equations\footnote{As a check, we are able to reproduce equation (15) of Zaqarashvili et al. (2007) using our approach.}.  We keep only first order terms in the Coriolis parameter, eliminate $b_{x_0}$ and $b_{y_0}$, and construct the $\hat{A}$ matrix:
\[ \hat{A} = \left( \begin{array}{ccc}
k_y \left[\omega\ - \frac{1}{\omega}\left(\frac{v_A}{H}\right)^2 \right] & -\left(\beta + i k_y f_0 \right) & -g k_x k_y\\
f_0 & i\left[ \left(\frac{v_A}{H} \right)^2\frac{1}{\omega} - \omega \right] & ig k_y\\
k_x H & k_y H & -\omega \end{array} \right).\]
As before, setting det$\hat{A}=0$ yields the dispersion relation,
\begin{equation}
\begin{split}
&\omega^4 - \omega^2 \left[ gHk^2 + f^2_0  + 2\left(v_A/H\right)^2 \right] \\
&- gHk_x\beta\omega + (v_A/H)^2\left[gHk^2 + (v_A/H)^2 \right] = 0,\\
\end{split}
\label{eq:dispersion_betaplane_mhd}
\end{equation}
which reduces to equation (\ref{eq:dispersion_fplane_mhd}) when $\beta=0$ as expected, i.e., the $\beta$-plane approximation reduces to the $f$-plane one at the poles.  For very slow waves ($\omega \ll f$) with large wavelengths ($kH \ll 1$), i.e., magneto-Rossby waves, the phase speed is
\begin{equation}
c_{\rm R} \approx - \frac{1}{2}\left[ {\cal C}_1 + \sqrt{ {\cal C}^2_1 + \left(\frac{\lambda_{\rm R}}{\lambda_{\rm B}}\right)^2 \left(\frac{\lambda_x}{\lambda_{\rm B}}\right)^2 c^2_0 {\cal C}^{-1}_2 } \right],
\end{equation}
where
\begin{equation}
\begin{split}
&{\cal C}_1 \equiv \beta \lambda^2_{\rm R} /{\cal C}_2,\\
&{\cal C}_2 \equiv \sin^2{\Theta} + \left(\lambda_{\rm R}/\lambda_{\rm B} \right)^2,\\
&\lambda_x \equiv 1/k_x.\\
\end{split}
\end{equation}
The negative sign implies that the phase propagation is westward. The root of the dispersion relation with the slowest angular frequency has the phase speed:
\begin{equation}
c_{p_x} \approx \frac{1}{2}\left[ \sqrt{ {\cal C}^2_1 + \left(\frac{\lambda_{\rm R}}{\lambda_{\rm B}}\right)^2 \left(\frac{\lambda_x}{\lambda_{\rm B}}\right)^2 c^2_0 {\cal C}^{-1}_2 } - {\cal C}_1 \right].
\end{equation}
 When $B_z \rightarrow 0$ ($\lambda_{\rm B} \rightarrow \infty$), we have $c_{p_x} \rightarrow 0$.  When $\beta=0$ and $\lambda_{\rm R}/\lambda_{\rm B} \ll 1$, we have $c_{p_x} \ll c_0$. This root can plausibly be associated with the east magnetostrophic mode. However, when considering the equatorially-confined, non-planar eigenfunctions on the $\beta$-plane (see Appendix \ref{append:betaplane}), we do not find this mode, yet do recover the two magneto-Poincar\'{e} and magneto-Rossby modes. This suggests that the magnetostrophic modes do not reside near the equator on the sphere, and the $\beta$-plane treatment may be inconclusive for these modes. We will confirm this in the next section. 

\section{MHD Shallow Water Waves: Spherical Coordinates}
\label{sect:mhd_spherical}

\subsection{Equations}

By analogy with equation (\ref{eq:b_gradient_approx}), we specify a radial magnetic field and allow for perturbations in the $\theta-$ and $\phi$-directions ($b_\theta$ and $b_\phi$, respectively).  The linearized equations of motion become:
\begin{equation}
\begin{split}
&\frac{\partial \hat{v}_\theta}{\partial t} - 2\Omega \hat{v}_\phi \cos{\theta} + \frac{g}{R} \sin{\theta} \frac{\partial h^\prime}{\partial \theta} + \frac{B_r \hat{b}_\theta}{4 \pi \rho H}= 0,\\
&\frac{\partial \hat{v}_\phi}{\partial t} + 2\Omega \hat{v}_\theta \cos{\theta} + \frac{g}{R} \frac{\partial h^\prime}{\partial \phi} + \frac{B_r \hat{b}_\phi}{4 \pi \rho H}= 0,\\
&\left(1 - \mu^2 \right) \frac{\partial h^\prime}{\partial t} + \frac{H\sin{\theta}}{R} \frac{\partial \hat{v}_\theta}{\partial \theta}  + \frac{H}{R} \frac{\partial \hat{v}_\phi}{\partial \phi} = 0,\\
&\frac{\partial \hat{b}_\theta}{\partial t} = \frac{\hat{v}_\theta B_r}{H},\\
&\frac{\partial \hat{b}_\phi}{\partial t} = \frac{\hat{v}_\phi B_r}{H},\\
\end{split}
\label{eq:linear_spherical_mhd}
\end{equation}
where $R$ is the radius of the sphere and $\theta = 90^\circ - \Theta$ is the co-latitude.  We define the following quantities\footnote{One of the earliest papers to use at least some of these transformations is Margules (1893).}:
\begin{equation}
\begin{split}
&\mu \equiv \cos{\theta},\\
&\hat{D} \equiv -\sin{\theta} \frac{\partial}{\partial \theta} = \left(1 - \mu^2 \right) \frac{\partial}{\partial \mu},\\
&\varpi \equiv \frac{\omega}{2\Omega},\\
&\epsilon \equiv \frac{4\Omega^2 R^2}{gH} = \left(\frac{R}{\lambda_{\rm R}}\right)^2,\\
&\hat{v}_\theta \equiv v^\prime_\theta \sin{\theta},\\
&\hat{v}_\phi \equiv v^\prime_\phi \sin{\theta},\\
&\hat{b}_\theta \equiv b_\theta \sin{\theta},\\
&\hat{b}_\phi \equiv b_\phi \sin{\theta}.\\
\end{split}
\label{eq:spherical_transformations}
\end{equation}
The introduction of $\hat{v}_\theta$ and $\hat{v}_\phi$ allows one to avoid singularities associated with $\theta=0^\circ$.

We seek the wave solutions,
\begin{equation}
\begin{split}
&\hat{v}_\theta = v_{\theta_0} \exp{i\left(s\phi - \omega t \right)},\\
&\hat{v}_\phi = v_{\phi_0} \exp{i\left(s\phi - \omega t \right)},\\
&h^\prime = h_0 \exp{i\left(s\phi - \omega t \right)},\\
\end{split}
\end{equation}
and eliminate the amplitudes for $\hat{b}_\theta$ and $\hat{b}_\phi$ to obtain:
\begin{equation}
\begin{split}
&\chi \tilde{v}_{\theta_0} + \mu v_{\phi_0} + \hat{D} \eta_0 = 0, \\
&\mu \tilde{v}_{\theta_0} + \chi v_{\phi_0} - s \eta_0 = 0, \\
&\varpi \epsilon \left( 1 - \mu^2 \right) \eta_0 - \hat{D}\tilde{v}_{\theta_0} - s v_{\phi_0}  = 0,\\
\end{split}
\label{eq:amplitudes_spherical_mhd}
\end{equation}
where $s$ is the toroidal wavenumber, $\tilde{v}_{\theta_0} \equiv i v_{\theta_0}$ and $\eta_0 \equiv gh_0/2\Omega R$.  Notice that the spherical hydrodynamic amplitude equations [equation (\ref{eq:amplitudes_spherical})] and equation (\ref{eq:amplitudes_spherical_mhd}) have identical structures, except that
\begin{equation}
\chi \equiv \varpi - \frac{\varpi^2_A}{\varpi}
\label{eq:chi}
\end{equation}
takes the place of $\varpi$ in a couple of places, with $v_A \equiv B_r/2\sqrt{\pi\rho}$ being the Alfv\'{e}n speed and 
\begin{equation}
\varpi_A \equiv \frac{v_A}{2\Omega H} = \frac{\lambda_{\rm R}}{\sqrt{2} \lambda_{\rm B}}.
\end{equation}

The equation for $\tilde{v}_{\theta_0}$, obtained from the set (\ref{eq:amplitudes_spherical_mhd}), has the form 
\begin{equation}
\hat{L}_s \tilde{v}_{\theta_0} = 0,
\label{eq:fundamental_eqn}
\end{equation}
where the operator is
\begin{equation}
\begin{split}
&\hat{L}_s \equiv \frac{d}{d \mu}\left[\left( 1 - \mu^2 \right) \frac{d}{d\mu} \right] - \frac{s^2}{1 - \mu^2} - \frac{s}{\chi} + \tilde{\epsilon}\left(\chi^2 - \mu^2 \right)\\
& - \frac{2 \tilde{\epsilon} \mu \chi \left(\chi \hat{D} - s\mu \right)}{s^2 - \tilde{\epsilon} \chi^2 \left( 1 - \mu^2 \right)}.\\
\end{split}
\label{eq:fundamental_eqn_operator_mhd}
\end{equation}
In the place of $\epsilon$, we now have $\tilde{\epsilon} \equiv \varpi\epsilon /\chi$ in the MHD case.

\subsection{Near-Equator Solutions ($\tilde{\epsilon} \gg 1$)}
\label{subsect:mhd_spherical_equator}

When $\tilde{\epsilon} \gg 1$, the governing equation for $\tilde{v}_{\theta_0}$ reduces to the spheroidal wave equation with the separation constant,
\begin{equation}
\Lambda_{sn}\left(\tilde{q} \right) = \tilde{\epsilon} \chi^2 - \frac{s}{\chi},
\label{eq:separation2}
\end{equation}
where $\tilde{q}=\sqrt{\tilde{\epsilon}}$.  The dispersion relation is
\begin{equation}
\begin{split}
&\varpi^8 - 4\varpi^2_A \varpi^6 - \left(\frac{2s}{\epsilon}\right) \varpi^5 + \left[6 \varpi^4_A - \frac{\left(2l+1 \right)^2}{\epsilon} \right] \varpi^4,\\
&+ \left(\frac{4s \varpi^2_A}{\epsilon}\right) \varpi^3 + \left[\frac{\varpi^2_A \left(2l+1 \right)^2}{\epsilon} - 4\varpi^6_A + \left(\frac{s}{\epsilon}\right)^2 \right] \varpi^2.\\
&- \left(\frac{2s \varpi^4_A}{\epsilon}\right) \varpi + \varpi^8_A = 0.\\
\end{split}
\label{eq:dispersion_spherical_mhd_equator}
\end{equation}
Eight roots exist for equation (\ref{eq:dispersion_spherical_mhd_equator}), but only three of these correspond to the angular frequencies for the magneto-Rossby and magneto-Poincar\'{e} modes.  We checked this by  requiring the eigenfrequencies and their corresponding eigenfunctions, which will be derived shortly, to satisfy the set of equations in (\ref{eq:linear_spherical_mhd}) in the appropriate limits.

We define
\begin{equation}
\begin{split}
&\tilde{X} \equiv \tilde{\epsilon}^{1/4} \mu,\\
&\tilde{\Psi} \equiv \exp{\left(-\frac{\tilde{X}^2}{2} \right)} ~\exp{i \left(s\phi - \omega t \right)}.\\
\end{split}
\end{equation}
Notice that as the magnetic field strength increases, the waves are ``pinched'' closer to the equator because of the the $\exp{(-\tilde{X}^2/2)}$ term.

Near the equator ($\vert \mu \vert \ll 1$), the spheroidal wave equation reduces to the parabolic cylinder equation.  Its solution is
\begin{equation}
v^\prime_{\theta} \approx -i \tilde{\eta}_0 ~\tilde{{\cal H}}_l ~\tilde{\Psi},
\end{equation}
where $\tilde{{\cal H}}_l \equiv {\cal H}_l(\tilde{X})$, ${\cal H}_l(\tilde{X})$ is the Hermite polynomial, $l \equiv n - s$ and $n$ plays the role of the poloidal wave number.  The quantity $\tilde{\eta}_0 = g \tilde{h}_0 / 2\Omega R$ is determined once the normalization $\tilde{h}_0$ is specified.  Using the amplitude equations in (\ref{eq:amplitudes_spherical_mhd}), we have:
\begin{equation}
\begin{split}
&\eta_0 \approx \frac{\left( \chi \hat{D} - s\mu \right) \tilde{v}_{\theta_0}}{\tilde{\epsilon} \chi^2 - s^2},\\
&v_{\phi_0} = \frac{\left(s \eta_0 - \mu \tilde{v}_{\theta_0} \right)}{\chi}.\\
\end{split}
\end{equation}
Also,
\begin{equation}
\hat{D} \tilde{v}_{\theta_0} \approx \tilde{\eta}_0 ~\tilde{\epsilon}^{1/4} ~\left( l\tilde{{\cal H}}_{l-1} - \frac{1}{2} \tilde{{\cal H}}_{l+1} \right) ~\tilde{\Psi}.
\end{equation}

The west magneto-Poincar\'{e} and the magneto-Rossby modes share the same eigenfunctions:
\begin{equation}
\begin{split}
&b_{\theta} \approx \frac{B_r}{\omega H} ~\tilde{\eta}_0 ~\tilde{{\cal H}}_l ~\tilde{\Psi},\\
&v^\prime_{\phi} \approx -\frac{s \tilde{\eta}_0}{\left(\tilde{\epsilon} \chi^2 - s^2\right) \chi} \left( l \tilde{{\cal H}}_{l-1} {\cal B}_+ + \frac{1}{2} \tilde{{\cal H}}_{l+1} {\cal B}_- \right) ~\tilde{\Psi},\\
&b_{\phi} \approx  -i \frac{B_r}{\omega H} \frac{s \tilde{\eta}_0}{\left(\tilde{\epsilon} \chi^2 - s^2\right) \chi} \left( l \tilde{{\cal H}}_{l-1} {\cal B}_+ + \frac{1}{2} \tilde{{\cal H}}_{l+1} {\cal B}_- \right) ~\tilde{\Psi},\\
&h^\prime \approx -\frac{\tilde{h}_0}{\tilde{\epsilon} \chi^2 - s^2} ~\left( l \tilde{{\cal H}}_{l-1} {\cal A}_+ + \frac{1}{2} \tilde{{\cal H}}_{l+1} {\cal A}_- \right) ~\tilde{\Psi},\\
\end{split}
\label{eq:mhd_spherical_eigenfunctions_west}
\end{equation}
where we have
\begin{equation}
\begin{split}
&{\cal A}_\pm \equiv s \tilde{\epsilon}^{-1/4} \pm \tilde{\epsilon}^{1/4} \chi,\\
&{\cal B}_\pm \equiv {\cal A}_\pm + \frac{\chi^2 \tilde{\epsilon} - s^2}{s} \tilde{\epsilon}^{-1/4}.\\
\end{split}
\end{equation}

The east magneto-Poincar\'{e} mode is described by the solutions:
\begin{equation}
\begin{split}
&b_{\theta} \approx \frac{B_r}{\omega H} ~\tilde{\eta}_0 ~\tilde{{\cal H}}_l ~\tilde{\Psi},\\
&v^\prime_{\phi} \approx -\frac{s \tilde{\eta}_0}{\left(\tilde{\epsilon} \chi^2 - s^2\right) \chi} \left( l \tilde{{\cal H}}_{l-1} {\cal B}_- + \frac{1}{2} \tilde{{\cal H}}_{l+1} {\cal B}_+ \right) ~\tilde{\Psi},\\
&b_{\phi} \approx  -i \frac{B_r}{\omega H} \frac{s \tilde{\eta}_0}{\left(\tilde{\epsilon} \chi^2 - s^2\right) \chi} \left( l \tilde{{\cal H}}_{l-1} {\cal B}_- + \frac{1}{2} \tilde{{\cal H}}_{l+1} {\cal B}_+ \right) ~\tilde{\Psi},\\
&h^\prime \approx -\frac{\tilde{h}_0}{\tilde{\epsilon} \chi^2 - s^2} ~\left( l \tilde{{\cal H}}_{l-1} {\cal A}_- + \frac{1}{2} \tilde{{\cal H}}_{l+1} {\cal A}_+ \right) ~\tilde{\Psi},\\
\end{split}
\label{eq:mhd_spherical_eigenfunctions_east}
\end{equation}

It is worth noting that the equatorial eigenfunctions can be obtained in the $\beta$-plane approximation as well (see Appendix \ref{append:betaplane}).

\subsection{Near-Pole Solutions ($-\tilde{\epsilon} \gg 1$)}
\label{subsect:mhd_spherical_pole}

Near the poles, the governing equation for the latitudinal velocity perturbation is Whittaker's equation (see Appendix \ref{subsect:classical_spherical_pole}).  The $\hat{L}_s$ operator in the hydrodynamic [equation (\ref{eq:fundamental_eqn_operator})] and MHD [equation (\ref{eq:fundamental_eqn_operator_mhd})] cases are mathematically identical, except that $\tilde{\epsilon}$ and $\chi$ take the places of $\epsilon$ and $\varpi$, respectively.  Therefore, the analysis performed in Appendix \ref{subsect:classical_spherical_pole} can be identically applied here.  However, the $\chi \approx -1$ solution yields the {\it east} magnetostrophic mode, because
\begin{equation}
\chi \approx -1 \Longrightarrow \varpi \approx \frac{1}{2} \left(-1 + \sqrt{1 + 4 \varpi^2_A} \right).
\end{equation}
Likewise, the $\chi \approx +1$ solution yields the {\it west} magnetostrophic mode.  We assume the form
\begin{equation}
\chi = \pm 1 + \frac{Q}{\sqrt{-\tilde{\epsilon}}},
\end{equation}
from which the dispersion relation for the slow modes is
\begin{equation}
\begin{split}
&\varpi^4 \mp 2\varpi^3 + \varpi^2 \left(1 + \frac{Q^2}{\epsilon} - 2\varpi^2_A \right) \\
&\pm 2\varpi^2_A \varpi + \varpi^2_A \left(\varpi^2_A - \frac{Q^2}{\epsilon} \right) = 0,\\
\end{split}
\label{eq:dispersion_spherical_mhd_pole}
\end{equation}
where $Q = 2m + 2l + 1$; for $\chi \approx \pm 1$, we have $m = \vert s \pm 1 \vert$.

The solution to equation (\ref{eq:fundamental_eqn}) is
\begin{equation}
\tilde{v}_{\theta_0} \approx V_0 ~\exp{\left(-\frac{\tilde{Y}}{2}\right)} ~\tilde{Y}^{\left(m + 1 \right)/2} ~{\cal L}^{\left( m \right)}_l \left(\tilde{Y}\right),
\end{equation}
where $V_0$ is an arbitrary normalization constant, ${\cal L}^{\left( m \right)}_l(\tilde{Y})$ is the associated Laguerre polynomial and
\begin{equation}
\tilde{Y} \equiv \sqrt{-\tilde{\epsilon}} \left( 1 - \mu^2 \right).
\end{equation}
The other amplitudes can be computed using:
\begin{equation}
\begin{split}
&\eta_0 = \frac{\left( \chi \hat{D} - s\mu \right) \tilde{v}_{\theta_0}}{\tilde{\epsilon} \chi^2 \left(1-\mu^2\right) - s^2},\\
&v_{\phi_0} = \frac{\left(s \eta_0 - \mu \tilde{v}_{\theta_0} \right)}{\chi}.\\
\end{split}
\end{equation}
Knowledge of the amplitudes allows one to compute the magnetic field perturbations by taking the real parts of the following expressions:
\begin{equation}
\begin{split}
&b_\theta = \frac{i B_r v^\prime_\theta}{\omega H},\\
&b_\phi = \frac{i B_r v^\prime_\phi}{\omega H}.\\
\end{split}
\end{equation}
Only two of the roots in equation (\ref{eq:dispersion_spherical_mhd_pole}) are physical; we checked this by again requiring the eigenfunctions to satisfy the set of equations in (\ref{eq:linear_spherical_mhd}) in the appropriate limits.

\begin{figure}
\begin{center}
\includegraphics[width=3.5in]{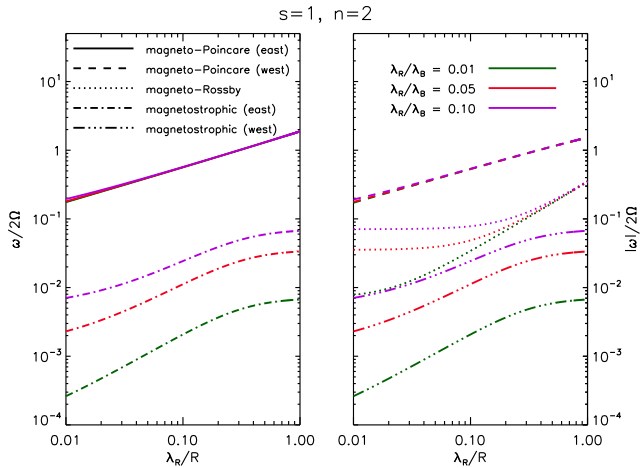}
\end{center}
\caption{Solutions to the dispersion relations for MHD shallow water wave systems in spherical geometry ($s=1$ and $n=2$).  Note that for the magneto-Poincar\'{e} modes, the curves for the three different values of $\lambda_{\rm R}/\lambda_{\rm B}=0.01, 0.05$ and $0.1$ overlap.}
\label{fig:w}
\end{figure}

\section{Discussion}
\label{sect:discussion}

\subsection{Existence of Wave Modes}

Our analyses in the previous sections have shown that five wave modes exist in MHD shallow water systems.  We now discuss a more intuitive way of understanding why they exist.  Firstly, consider fluid flow on the surface of a non-rotating, non-magnetized cylinder with gravity.  Only gravity waves exist and the flow can generally be eastward- or westward-propagating.  The Rossby mode does not exist even if we rotate the cylinder (hence producing the Poincar\'{e} modes), because it requires the presence of a latitudinally-varying Coriolis force.

If we now allow for the presence of a magnetic field, the magnetostrophic modes appear alongside the magneto-Poincar\'{e} modes.  A magnetized, self-gravitating cylinder --- regardless of whether it is rotating --- possesses four shallow water wave modes.  If we replace the cylinder with a sphere, the magneto-Rossby mode appears.  Shallow water systems on a non-magnetized sphere only have three modes: the Poincar\'{e} and Rossby modes.

Note that unlike on the $\beta$-plane, there are both eastward- and westward-propagating magnetostrophic waves on a magnetized sphere. As these waves are more concentrated towards the poles, they are less susceptible to the effects of varying the Coriolis parameter over the sphere.  Furthermore --- unlike in the case of the hydrodynamic Rossby mode --- the magneto-Rossby mode exists even for $\beta \rightarrow 0$, where it then reduces to the west magnetostrophic mode.

\begin{figure}
\begin{center}
\includegraphics[width=3.2in]{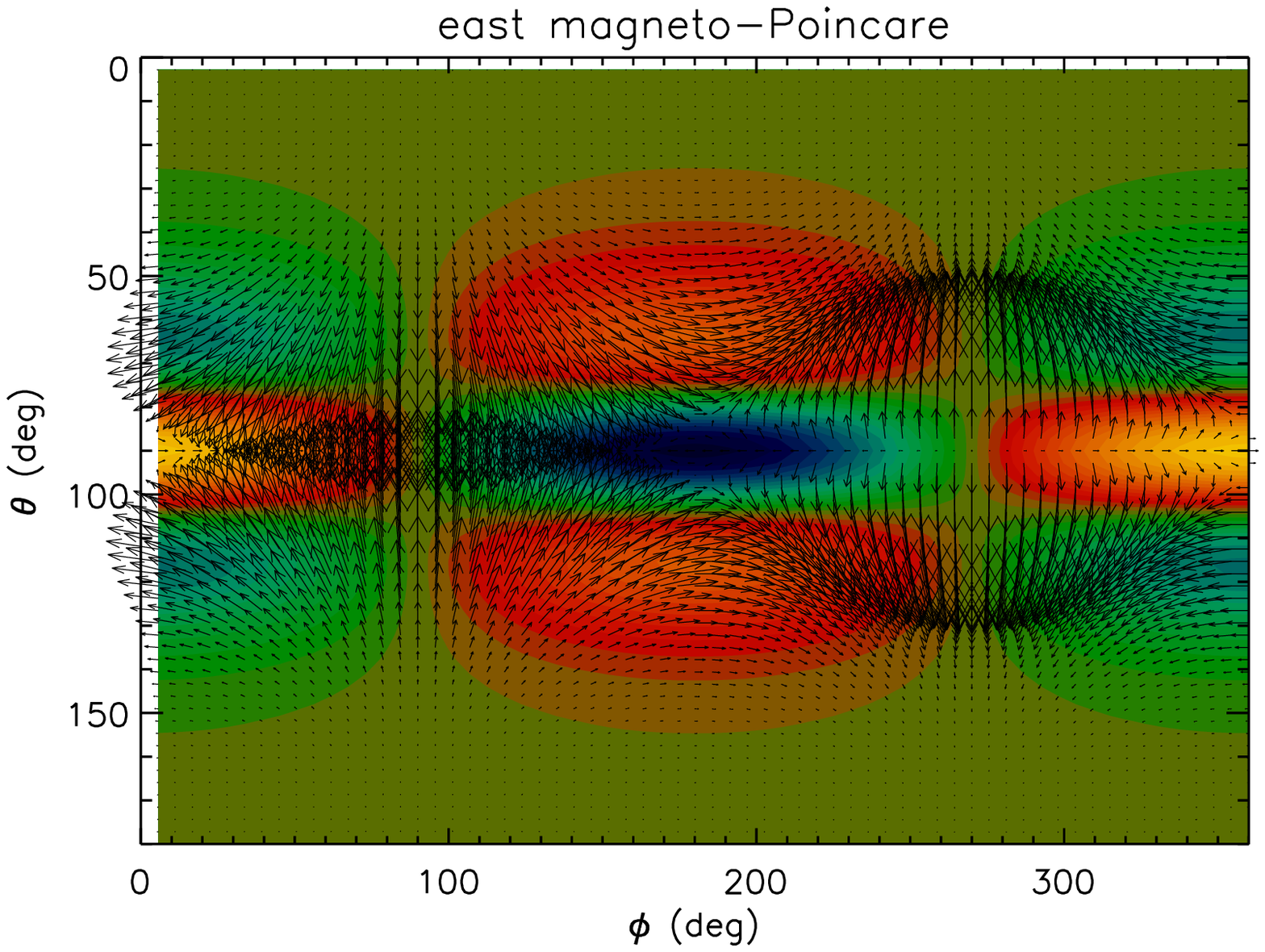}
\includegraphics[width=3.2in]{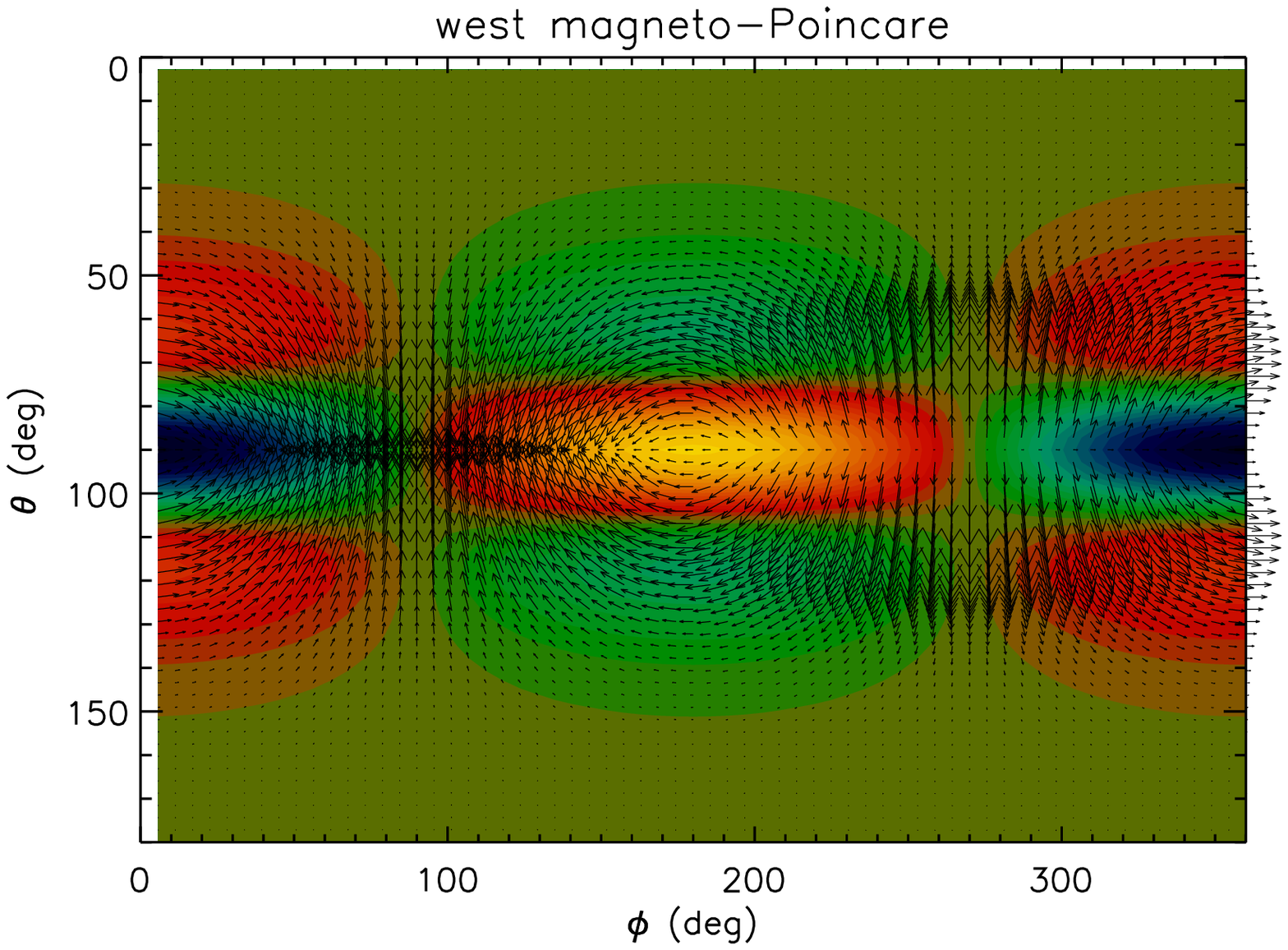}
\includegraphics[width=3.2in]{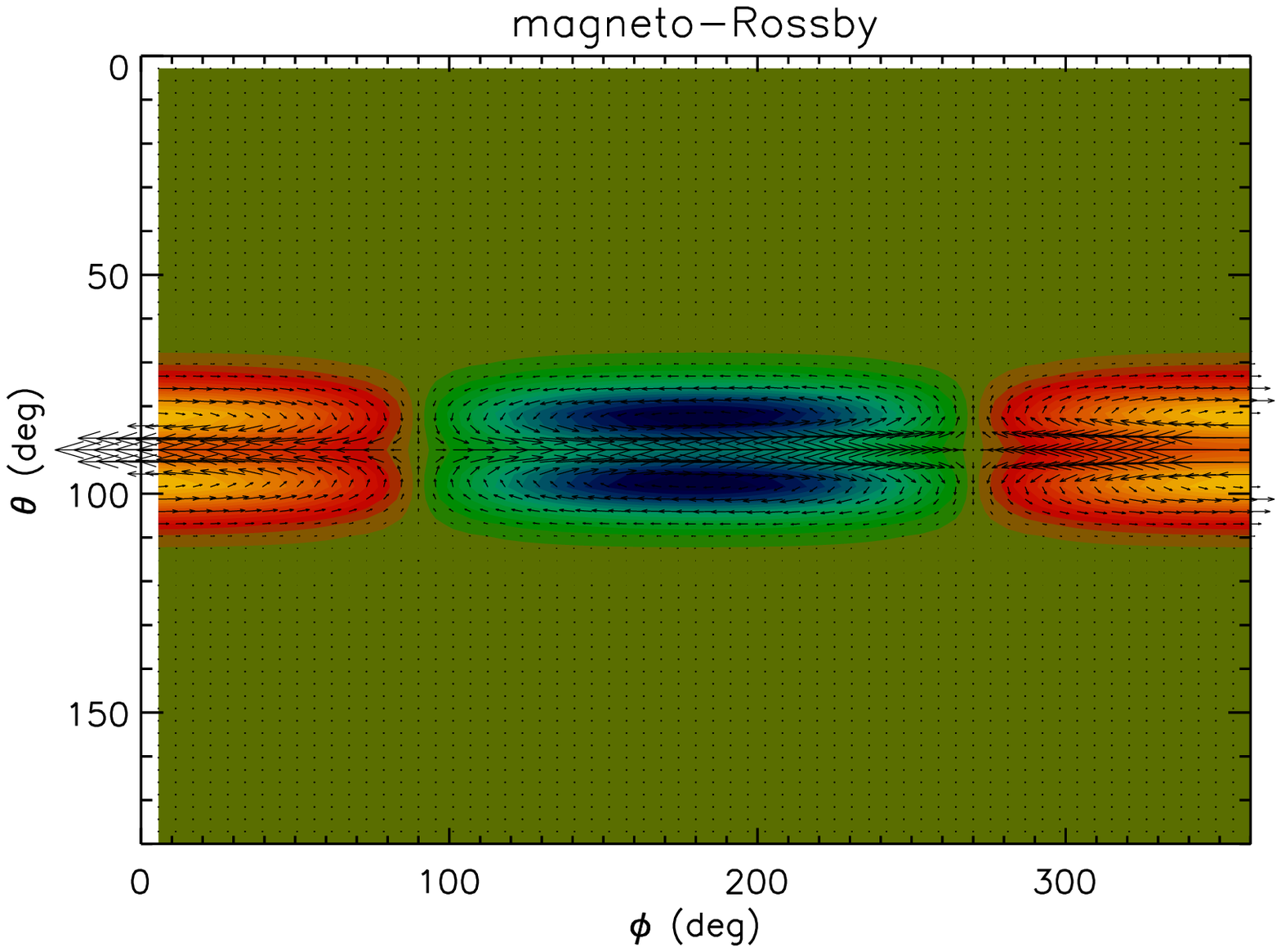}
\end{center}
\caption{Wave solutions ($s=1$ and $n=2$) for the magneto-Poincar\'{e} and magneto-Rossby modes.  For illustration, we have adopted $B_r = 10^8$ G, $H=10^2$ cm, $\rho = 10^6$ g cm$^{-3}$, $\Omega = 10^3$ rad s$^{-1}$, $g=2 \times 10^{14}$ cm s$^{-2}$ and $R=10$ km.  The contours represent the ocean height with light and dark colours corresponding to positive and negative perturbations, respectively; the height normalization is arbitrary.  The arrows represent the velocity field.}
\label{fig:eigenfunctions_fast_rossby}
\end{figure}

\subsection{Angular Frequencies}

The dispersion relations for the MHD modes on a sphere, described by equations (\ref{eq:dispersion_spherical_mhd_equator}) and (\ref{eq:dispersion_spherical_mhd_pole}), are parametrized by the ratio of Rossby ($\lambda_{\rm R}$) to Alfv\'{e}n ($\lambda_{\rm B}$) radii,
\begin{equation}
\frac{\lambda_{\rm R}}{\lambda_{\rm B}} = 0.02 B_8 \rho^{-1/2}_6 H^{-1}_3 \Omega^{-1}_3,
\label{eq:length_ratio}
\end{equation}
where for illustration, we have adopted $B_8 = B_r/10^8$ G, $\rho_6 = \rho/10^6$ g cm$^{-3}$, $H_3 = H/10^3$ cm, $\Omega_3 = \Omega/10^3$ rad s$^{-1}$, $g=2 \times 10^{14}$ cm s$^{-2}$ and $R=10$ km, following Spitkovsky, Levin \& Ushomirsky (2002) in the context of neutron stars.  For these values, we have $\lambda_{\rm R} \approx 2$ km and $\lambda_{\rm B} \approx 100$ km ($v_A \approx 0.3$ km s$^{-1}$).

In Fig. \ref{fig:w}, we calculate the angular frequencies of the wave modes for various values of $\lambda_{\rm R}/\lambda_{\rm B}$.  Three features are apparent.  Firstly, the magneto-Rossby mode has an angular frequency intermediate between those of the magnetostrophic and magneto-Poincar\'{e} modes.  The magneto-Poincar\'{e} modes are insensitive to $\lambda_{\rm R}/\lambda_{\rm B}$.  However, the magnetostrophic modes are sensitive to $\lambda_{\rm R}/\lambda_{\rm B}$, because $\vert \varpi \vert \sim \varpi^2_A$.  Since $\varpi_A \propto B_r$, such a property can conceivably be used as a magnetometer, provided the other quantities are known.

\subsection{Visualization of the Wave Modes}

\begin{figure}
\begin{center}
\includegraphics[width=3.2in]{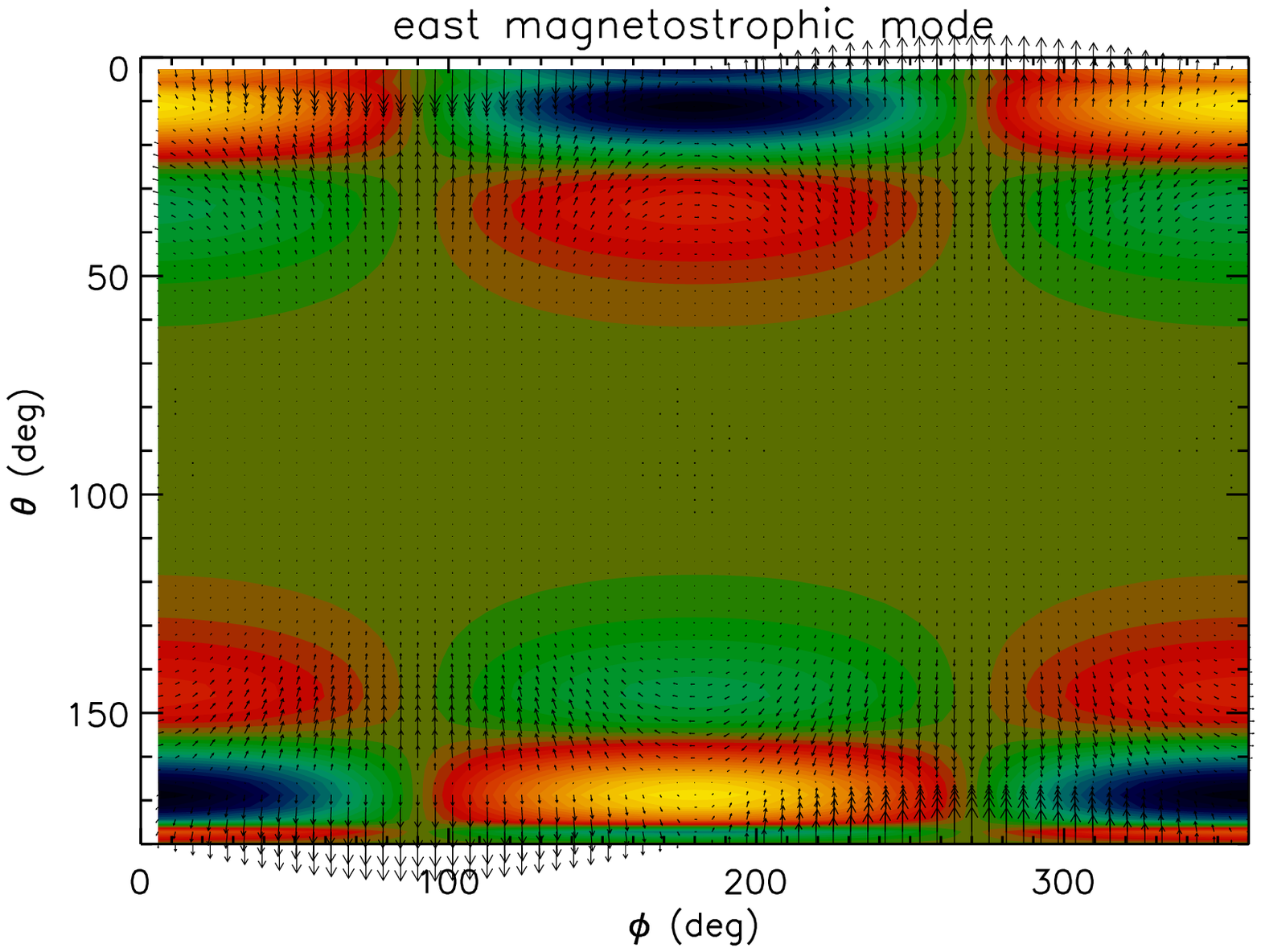}
\includegraphics[width=3.2in]{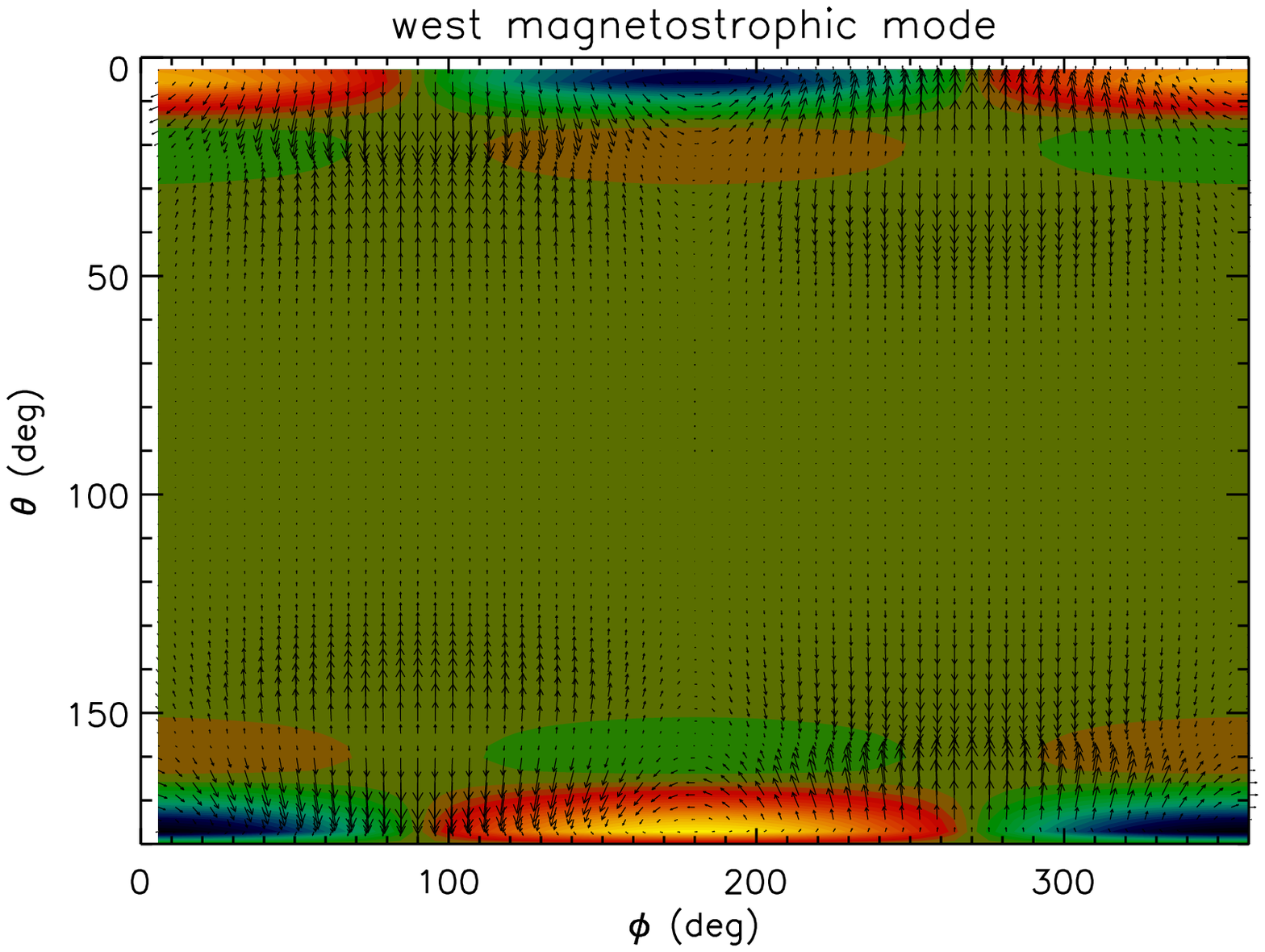}
\end{center}
\caption{Same as Fig. \ref{fig:eigenfunctions_fast_rossby}, but for the magnetostrophic modes with $B_r = 10^9$ G.}
\label{fig:eigenfunctions_slow}
\end{figure}

In Figs. \ref{fig:eigenfunctions_fast_rossby} and \ref{fig:eigenfunctions_slow}, we show two-dimensional plots of the MHD wave solutions for the magneto-Poincar\'{e}, magneto-Rossby and magnetostrophic modes.  We adopt $s=1$ and $n=2$ for illustration, but note that the toroidal ($s$) and poloidal ($n$) wave numbers have geometrical interpretations, corresponding to $n - \vert s \vert +1$ and $2\vert s \vert$ wave nodes (i.e., ``zero crossings'') in the latitudinal and longitudinal directions, respectively.

The qualitative behaviour of the modes agrees with the description of Matsuno (1966).  In particular, the magneto-Rossby mode can be understood in the following manner: the height gradient --- and therefore the pressure gradient --- is positive from the low pressure region in the middle ($\phi = 180^\circ$) to the high pressure region on the right (Fig. \ref{fig:eigenfunctions_fast_rossby}).  Since $v^\prime_y \propto \partial h^\prime/\partial x$, we get $v^\prime_y > 0$ and fluid flowing {\it away from} the equator.  The reverse argument holds for the height gradient (because $dh^\prime > 0$ and $dx < 0$) between the low pressure region and the high pressure region on the left ($\phi = 0^\circ$), resulting in $v^\prime_y < 0$ and fluid flowing {\it towards} the equator.

A feature that is absent from hydrodynamic systems is the magnetic pinching of the waves.  Increasing $B_r$ concentrates the magneto-Rossby and magneto-Poincar\'{e} waves closer to the equator; the magnetostrophic waves are concentrated closer to the poles.  Although our solutions formally hold only in the large magnetic field limit, we verified the concentration of eigenfunctions and the values of eigenfrequencies for different modes by direct numerical integration of the spherical equations in (\ref{eq:linear_spherical_mhd}).  We used the method of Ivanov (2007) to find the regular solutions at the poles by series expansion and then applied the shooting method to determine the eigenfrequencies of the odd and even modes. 

\subsection{Type I X-ray Bursts from Neutron Stars}

Type I X-ray bursts are non-catastrophic, thermonuclear explosions occurring on accreting neutron stars in low-mass X-ray binaries.  They have typical rise and decay times $\sim 1$ s and $\sim 10$---100 s, respectively (e.g., Strohmayer \& Bildsten 2006), and are often accompanied by millisecond ($\sim 300$---600 Hz) oscillations (see van der Klis [2000] for a review).  As the X-ray burst reaches its peak luminosity and fades off, oscillations with frequency drifts $\sim 1$ to 5 Hz, relative to some asymptotic frequency, persist for a time $\sim 10$ s.  The asymptotic frequency is usually identified with the rotational frequency of the neutron star (e.g., Strohmayer et al. 1998).

A popular interpretation is that these oscillations are caused by the rotational modulation of a growing ``hot spot'' on the surface of the neutron star.  The timescale for thermonuclear burning is much shorter than the time needed to accumulate enough fuel to trigger thermal instabilities, allowing localized ignitions on the stellar surface.  Initially, the hot spot moves backwards against the sense of rotation due to the (approximate) conservation of potential vorticity, much like westward-moving storms on Earth; it then spreads and engulfs the entire star (Spitkovsky, Levin \& Ushomirsky 2002).  Transverse pressure gradients cannot be ignored because the ignited ocean of material is not fully degenerate (Cumming \& Bildsten 2000), implying that an important characteristic velocity is the shallow water wave speed.

Previous studies (e.g., Heyl 2004; Berkhout \& Levin 2008) have dealt with purely hydrodynamic shallow water systems and have focused on the Rossby mode (``r-mode'').  It has been proposed that the frequency drift is caused by the Rossby mode, but it was quickly realized that the computed drifts are somewhat larger than what is observed.  Our results in Fig. \ref{fig:w} are consistent with this view: for  $\lambda_{\rm R}/\lambda_{\rm B} \sim 0.01$, the drift frequency is about $\nu = -\omega/2\pi \sim 100$ Hz (adopting the Spitkovsky, Levin \& Ushomirsky [2002] parameter values).  By contrast, the west magnetostrophic mode has the frequency (in the limit that $\epsilon \gg 1$)
\begin{equation}
\nu \sim \frac{\Omega}{4\pi} \left(\frac{\lambda_{\rm R}}{\lambda_{\rm B}} \right)^2 \propto \frac{B^2_r}{\rho \Omega H^2},
\end{equation}
which is of the correct order of magnitude (i.e., $\sim 1$ Hz) if $\lambda_{\rm R}/\lambda_{\rm B} \sim 0.1$ and $\Omega \sim 10^3$ rad s$^{-1}$.  However, if we assume $B_r$ and $\rho$ to be constant with time, then as the accreted material cools and $H$ decreases, $\nu$ increases, contrary to what is required for explaining the temporal behaviour of the frequency drifts.  Clearly, a full explanation of the frequency drifts also involves understanding the physics of accreting material onto the surface of a neutron star (Inogamov \& Sunyaev 1999).

It is not the intention of our present study to compute detailed models for the frequency drift.  Rather, it is to point out that the west magnetostrophic mode is a promising candidate that deserves further attention.  Note that the analytical eigenfunctions derived in \S\ref{subsect:mhd_spherical_pole} are good approximations only for $B_r \gtrsim 10^9$ G (with $H \sim 10^2$---$10^3$ cm), whereas Type I X-ray bursters are believed to be $\sim 10^7$---$10^8$ G systems.

\subsection{Future Work}

Our study of MHD shallow water waves is generic enough that it is readily amenable to various improvements.  Natural extensions will be to consider viscous fluids with more general magnetic field configurations.  The linear analysis can be generalized to a non-linear one via study of the Korteweg-deVries equation (e.g., KC04).

\section{Summary}
\label{sect:summary}

The salient points of our study are summarized as follows:
\begin{enumerate}

\item We have performed a linear analysis of inviscid, incompressible MHD waves in the shallow water approximation.  A generic feature of such systems is the existence of five wave modes: east and west magneto-Poincar\'{e}; east and west magnetostrophic; and magneto-Rossby.

\item Analytical functions for the velocity, height and magnetic field perturbations are obtained in the limit that $\vert \tilde{\epsilon} \vert \gg 1$.  These functions are useful as a guide towards obtaining numerical solutions or performing simulations.

\item For reasonable values of the parameters, the magneto-Rossby and magneto-Poincar\'{e} modes belong to the $\tilde{\epsilon} \gg 1$ family of solutions and reside near the equator.  The magnetostrophic modes belong to the $-\tilde{\epsilon} \gg 1$ family of solutions and reside near the poles.  Increasing the magnetic field strength further concentrates the waves near the equator and poles, respectively.

\item The west magnetostrophic mode is a potential candidate for explaining the observed frequency drifts in Type I X-ray bursts from neutron stars.

\end{enumerate}

\acknowledgments
\scriptsize
K.H. acknowledges generous financial, secretarial and logistical support by the Institute for Advanced Study and by NSF grant AST-0807444.  He is grateful to Lim Hock for introducing him to shallow water waves, and benefited from useful conversations with Jonathan Mitchell, Yuri Levin, Martin Pessah, Phil Chang, Chris Thompson, Roger Blandford, Dick McCray and Shane Davis. A.S. acknowledges support from the Alfred P. Sloan Foundation
fellowship and NSF grant AST-0807381.
\normalsize

\appendix

\section{A. Classical Shallow Water Waves: Cartesian Coordinates}
\label{sect:classical_cartesian}

\subsection{A.1. Without Rotation (Gravity Waves)}
\label{subsect:gravity}

Hydrodynamic shallow water waves are described by Euler's equation:
\begin{equation}
\frac{\partial \vec{v}}{\partial t} + \vec{v}.\nabla \vec{v} = -\frac{1}{\rho} \nabla P.
\end{equation}
The fluid pressure is
\begin{equation}
P = P_0 + \rho g (h-z),
\end{equation}
where $P_0$ is constant, such that the $x$- and $y$-components of the Euler equations are
\begin{equation}
\begin{split}
&\frac{\partial v_x}{\partial t} + v_x \frac{\partial v_x}{\partial x} + v_y \frac{\partial v_x}{\partial y} = -g \frac{\partial h}{\partial x},\\
&\frac{\partial v_y}{\partial t} + v_x \frac{\partial v_y}{\partial x} + v_y \frac{\partial v_y}{\partial y} = -g \frac{\partial h}{\partial y}.\\
\end{split}
\label{eq:velocity_classical}
\end{equation}

An additional equation of motion comes from the incompressibility condition:
\begin{equation}
v_z = - \int^h_0 \left( \frac{\partial v_x}{\partial x} + \frac{\partial v_y}{\partial y} \right) dz = - \left( \frac{\partial v_x}{\partial x} + \frac{\partial v_y}{\partial y} \right)h.
\end{equation}
But we also have
\begin{equation}
v_z = \frac{Dh}{Dt} = \frac{\partial h}{\partial t} + v_x \frac{\partial h}{\partial x} + v_y \frac{\partial h}{\partial y}.
\end{equation}
It follows that
\begin{equation}
\frac{\partial h}{\partial t} + v_x \frac{\partial h}{\partial x} + v_y \frac{\partial h}{\partial y} + \left( \frac{\partial v_x}{\partial x} + \frac{\partial v_y}{\partial y} \right)h = 0,
\label{eq:height_classical}
\end{equation}
or, more compactly,
\begin{equation}
\frac{\partial h}{\partial t} + \nabla.\left(h \vec{v}\right) = 0.
\end{equation}
One can easily check that the derived equations in (\ref{eq:velocity_classical}) and (\ref{eq:height_classical}) satisfy ``state of calm'' conditions,
\begin{equation}
v_x=v_y=0, h=H_0,
\end{equation}
where $H_0$ is a constant.

Considering small perturbations to the velocity and height of the water waves yield the following linearized equations:
\begin{equation}
\begin{split}
&\frac{\partial v^\prime_x}{\partial t} + V_x \frac{\partial v^\prime_x}{\partial x} + V_y \frac{\partial v^\prime_x}{\partial y} + g \frac{\partial h^\prime}{\partial x} =0,\\
&\frac{\partial v^\prime_y}{\partial t} + V_x \frac{\partial v^\prime_y}{\partial x} + V_y \frac{\partial v^\prime_y}{\partial y} + g \frac{\partial h^\prime}{\partial y} =0,\\
&\frac{\partial h^\prime}{\partial t} + V_x \frac{\partial h^\prime}{\partial x} + V_y \frac{\partial h^\prime}{\partial y} + \left( \frac{\partial v^\prime_x}{\partial x} + \frac{\partial v^\prime_y}{\partial y} \right)H = 0.\\
\end{split}
\label{eq:linear_classical}
\end{equation}

Seeking wave solutions to equation (\ref{eq:linear_classical}) yields:
\begin{equation}
\begin{split}
&v_{x_0} A_0 + h_0 g k_x = 0,\\
&v_{y_0} A_0 + h_0 g k_y = 0,\\
&v_{x_0} k_x H + v_{y_0} k_y H + h_0 A_0 = 0.\\
\end{split}
\label{eq:amplitudes_classical}
\end{equation}

The linear, simultaneous equations in (\ref{eq:amplitudes_classical}) can be arranged in the form (i.e., ``Cramer's Rule''; e.g., Arfken \& Weber 1995):
\[ \hat{A} \left( \begin{array}{c}
v_{x_0}\\
v_{y_0} \\
h_0 \end{array} \right) = 0,\] where
\[ \hat{A} = \left( \begin{array}{ccc}
A_0 & 0 & g k_x\\
0 & A_0 & g k_y\\
k_x H & k_y H & A_0 \end{array} \right).\] 
The solutions for $v_{x_0}$, $v_{y_0}$ and $h_0$ are non-trivial only if
\begin{equation}
\mbox{det}\hat{A} = 0.
\end{equation}
It follows that
\begin{equation}
A_0 \left[ A^2_0 - gH \left( k^2_x + k^2_y \right) \right] = 0,
\end{equation}
which yields the following dispersion relations:
\begin{equation}
\omega  =
\begin{cases}
k_x V_x + k_y V_y,\\
k_x V_x + k_y V_y \pm \sqrt{gH \left( k^2_x + k^2_y \right)}.\\
\end{cases}
\end{equation}

The first of these dispersion relations is trivial; it simply describes an advection wave.  Setting $V_x = V_y = 0$ and analyzing the $\omega \ge 0$ solutions, we obtain the $x$- and $y$-components of the phase velocity:
\begin{equation}
\begin{split}
&c_{p_x} = \omega/k_x = c_0 \sqrt{1 + k^2_y/k^2_x},\\
&c_{p_y} = \omega/k_y = c_0 \sqrt{1 + k^2_x/k^2_y},\\
\end{split}
\end{equation}
such that
\begin{equation}
c_p = \sqrt{c^2_{p_x} + c^2_{p_y}} = c_0 \sqrt{2 + k^2_x/k^2_y + k^2_y/k^2_x}.
\end{equation}
Similarly, the $x$- and $y$-components of the group velocity are:
\begin{equation}
\begin{split}
&c_{g_x} = \frac{\partial \omega}{\partial k_x} = \frac{k_x c_0}{\sqrt{k^2_x + k^2_y}},\\
&c_{g_y} = \frac{\partial \omega}{\partial k_y} = \frac{k_y c_0}{\sqrt{k^2_x + k^2_y}},\\
\end{split}
\end{equation}
and
\begin{equation}
c_g = \sqrt{c^2_{g_x} + c^2_{g_y}} = c_0.
\end{equation}
Note that $c_g < c_p$ and the amplitudes $v_{x_0}$, $v_{y_0}$ and $h_0$ are always in phase.

\subsection{A.2. With Constant Rotation: $f$-Plane Treatment (Poincar\'{e} Waves)}
\label{subsect:fplane}

In the case of non-zero rotation, the Euler equation in the rotating frame has a Coriolis force term:
\begin{equation}
\frac{\partial \vec{v}}{\partial t} + \vec{v}.\nabla \vec{v} = -\frac{1}{\rho} \nabla P - 2\left( \vec{\Omega} \times \vec{v} \right).
\end{equation}
Assuming that $\vec{\Omega} = (0,0,\Omega)$, the equations of motion become
\begin{equation}
\begin{split}
&\frac{\partial v_x}{\partial t} + v_x \frac{\partial v_x}{\partial x} + v_y \frac{\partial v_x}{\partial y} = -g \frac{\partial h}{\partial x} + f v_y,\\
&\frac{\partial v_y}{\partial t} + v_x \frac{\partial v_y}{\partial x} + v_y \frac{\partial v_y}{\partial y} = -g \frac{\partial h}{\partial y} - f v_x,\\
\end{split}
\label{eq:velocity_fplane}
\end{equation}
with equation (\ref{eq:height_classical}) remaining unchanged.  Such a treatment with the Coriolis parameter, $f = 2\Omega$, held constant is called the ``$f$-plane approximation'' (see G82 and KC04).  It is possible to Taylor-expand $f$ and keep the first order term, which will have a spatial dependence.  Such a treatment is known as the ``$\beta$-plane approximation'' and is examined in Appendix \ref{subsect:betaplane}.

We again consider small perturbations to the physical quantities and linearize the equations:
\begin{equation}
\begin{split}
&\frac{\partial v^\prime_x}{\partial t} + V_x \frac{\partial v^\prime_x}{\partial x} + V_y \frac{\partial v^\prime_x}{\partial y} + g \frac{\partial h^\prime}{\partial x} - 2 \Omega v^\prime_y =0,\\
&\frac{\partial v^\prime_y}{\partial t} + V_x \frac{\partial v^\prime_y}{\partial x} + V_y \frac{\partial v^\prime_y}{\partial y} + g \frac{\partial h^\prime}{\partial y} + 2 \Omega v^\prime_x = 0,\\
\end{split}
\label{eq:linear_fplane}
\end{equation}
where the linearized equation of motion for $h^\prime$ remains unchanged (see equation [\ref{eq:linear_classical}]).

The matrix $\hat{A}$ now becomes
\[ \hat{A} = \left( \begin{array}{ccc}
iA_0 & -2\Omega & ig k_x\\
2\Omega & iA_0 & ig k_y\\
k_x H & k_y H & A_0 \end{array} \right),\]
and the non-trivial dispersion relation is
\begin{equation}
\omega = k_x V_x + k_y V_y \pm \sqrt{gH \left( k^2_x + k^2_y \right) + 4\Omega^2}.
\end{equation}
Such a dispersion relation agrees with that described by Holton (2004; pg. 208).  Waves with this dispersion relation are often called ``Poincar\'{e} waves'' (pg. 196 of G82), while Poincar\'{e} waves with a finite horizontal boundary are termed ``Kelvin waves'' (pg. 615 of KC04)\footnote{Poincar\'{e} waves are often referred to as ``gravity waves'' or ``g-modes'' with rotational effects implied.}.

Setting $V_x = V_y = 0$ (i.e., solid body rotation), the phase velocity is
\begin{equation}
c_p = \sqrt{gH \left( 2 + k^2_x/k^2_y + k^2_y/k^2_x \right) + 4\Omega^2 \left(1/k^2_x + 1/k^2_y \right)},
\end{equation}
while the group velocity is
\begin{equation}
c_g = c_0 \left[ \frac{k^2}{k^2 + 4\Omega^2/gH} \right]^{1/2}.
\end{equation}

\subsection{A.3. With Constant Rotation: $\beta$-Plane Treatment (Poincar\'{e} and Rossby Waves)}
\label{subsect:betaplane}

Consider the Coriolis parameter to be
\begin{equation}
f = f_0 + \beta y,
\label{eq:coriolis_parameter_fplane}
\end{equation}
where $f_0 = 2\Omega \sin{\Theta}$ and $\Theta$ denotes the latitude on a sphere.  The parameter $\beta$ is given by
\begin{equation}
\beta = \frac{2\Omega}{R} \cos{\Theta},
\end{equation}
with $R$ being the radius of the sphere.  Such a treatment is known as the ``$\beta$-plane approximation'' (e.g., KC04), and is the first step towards considering curvature effects on a sphere.

The set of linearized equations in the $\beta$-plane approximation is:
\begin{equation}
\begin{split}
&\frac{\partial v^\prime_x}{\partial t} + g \frac{\partial h^\prime}{\partial x} - f v^\prime_y =0,\\
&\frac{\partial v^\prime_y}{\partial t} + g \frac{\partial h^\prime}{\partial y} + f v^\prime_x = 0,\\
&\frac{\partial h^\prime}{\partial t} + \left( \frac{\partial v^\prime_x}{\partial x} + \frac{\partial v^\prime_y}{\partial y} \right)H = 0.\\
\end{split}
\label{eq:linear_betaplane}
\end{equation}
Differentiating the second equation in (\ref{eq:linear_betaplane}) twice with respect to $t$ yields:
\begin{equation}
\frac{\partial^3 v^\prime_y}{\partial t^3} + f^2 \frac{\partial v^\prime_y}{\partial t} + fgH \frac{\partial}{\partial x} \left( \frac{\partial v^\prime_x}{\partial x} + \frac{\partial v^\prime_y}{\partial y} \right) = gH \frac{\partial^2}{\partial y \partial t} \left( \frac{\partial v^\prime_x}{\partial x} + \frac{\partial v^\prime_y}{\partial y} \right).
\label{eq:kundu1}
\end{equation}
Differentiating the first and second equations in (\ref{eq:linear_betaplane}) with respect to $y$ and $x$ respectively, and subtracting them from one another gives:
\begin{equation}
\frac{\partial}{\partial t} \left( \frac{\partial v^\prime_x}{\partial y} - \frac{\partial v^\prime_y}{\partial x} \right) - f \left( \frac{\partial v^\prime_x}{\partial x} + \frac{\partial v^\prime_y}{\partial y} \right) - \beta v^\prime_y = 0.
\label{eq:kundu2}
\end{equation}
Adding equations (\ref{eq:kundu1}) and (\ref{eq:kundu2}), with the latter operated on by $gH\frac{\partial}{\partial x}$, eliminates $v^\prime_x$:
\begin{equation}
\frac{\partial^3 v^\prime_y}{\partial t^3} + f^2 \frac{\partial v^\prime_y}{\partial t} - gH \frac{\partial}{\partial t} \left( \frac{\partial^2 v^\prime_y}{\partial x^2} + \frac{\partial^2 v^\prime_y}{\partial y^2} \right) -  \beta gH \frac{\partial v^\prime_y}{\partial x} = 0.
\label{eq:kundu3}
\end{equation}
Such an approach is described in KC04 (pg. 610); a generalized version of it was applied to MHD systems by Zaqarashvili et al. (2007).  From equation (\ref{eq:kundu3}), one can seek wave solutions as usual and obtain the dispersion relation.

There is a more convenient approach one can adopt to obtain the dispersion relation.  Instead of juggling differential operators, one differentiates the first equation in (\ref{eq:linear_betaplane}) with respect to $y$; the purpose of this step is to extract the first order terms in $f$.  Together with the other two equations in (\ref{eq:linear_betaplane}), wave solutions are then sought and the matrix $\hat{A}$ is once again constructed:
\[ \hat{A} = \left( \begin{array}{ccc}
\omega k_y & -\left( \beta + i k_y f_0 \right) & -g k_x k_y\\
f_0 & -i\omega & ig k_y\\
k_x H & k_y H & -\omega \end{array} \right),\]
Setting det$\hat{A}=0$ yields the dispersion relation:
\begin{equation}
\omega^3 - \omega\left( gHk^2 + f^2_0 \right) - gHk_x\beta = 0.
\label{eq:dispersion_classical_betaplane}
\end{equation}
Notice that equation (\ref{eq:dispersion_classical_betaplane}) is asymmetric with respect to $k_x$ and $k_y$, implying that wave motion is not isotropic in the horizontal direction.  Very slow ($\omega \ll f$), large-wavelength ($kH \ll 1$) waves  are called ``Rossby waves''.  They are sometimes called ``planetary waves'', and have an angular frequency of (see also pg. 634 of KC04)
\begin{equation}
\omega \approx -\frac{\beta k_x}{ k^2 + \left(f_0/c_0\right)^2}.
\label{eq:frequency_classical_rossby}
\end{equation}
For very large wavelengths, the {\it eastward} phase speed is
\begin{equation}
c_{\rm R} \approx - \beta \left( c_0/f_0 \right)^2.
\end{equation}
The negative sign shows that phase propagation is always {\it westward}.

\section{B. Classical Shallow Water Waves: Spherical Coordinates}
\label{sect:classical_spherical}

We revisit the classic work of LH68 as a prelude to \S\ref{sect:mhd_spherical}.

\subsection{B.1. Equations}

Consider waves with a velocity $\vec{v} = (v_r,v_\theta,v_\phi)$.  The linearized equations of motion have the form:
\begin{equation}
\begin{split}
&\frac{\partial v^\prime_\theta}{\partial t} - 2\Omega v^\prime_\phi \cos{\theta} + \frac{g}{R} \frac{\partial h^\prime}{\partial \theta} = 0,\\
&\frac{\partial v^\prime_\phi}{\partial t} + 2\Omega v^\prime_\theta \cos{\theta} + \frac{g}{R \sin{\theta}} \frac{\partial h^\prime}{\partial \phi} = 0,\\
&\frac{\partial h^\prime}{\partial t} + \frac{H}{R \sin{\theta}} \left[ \frac{\partial}{\partial \theta}\left( v^\prime_\theta \sin{\theta} \right) + \frac{\partial v^\prime_\phi}{\partial \phi} \right] = 0.\\
\end{split}
\label{eq:linear_spherical}
\end{equation}

The amplitude equations become:
\begin{equation}
\begin{split}
&\varpi \tilde{v}_{\theta_0} + \mu v_{\phi_0} + \hat{D} \eta_0 = 0, \\
&\mu \tilde{v}_{\theta_0} + \varpi v_{\phi_0} - s \eta_0 = 0, \\
&\varpi \epsilon \left( 1 - \mu^2 \right) \eta_0 - \hat{D}\tilde{v}_{\theta_0} - s v_{\phi_0}  = 0.\\
\end{split}
\label{eq:amplitudes_spherical}
\end{equation}
The second equation in the preceding set can be used to eliminate $v_{\phi_0}$ and obtain $\hat{L}_s \tilde{v}_{\theta_0} = 0$, where the operator is
\begin{equation}
\hat{L}_s \equiv \frac{d}{d \mu}\left[\left( 1 - \mu^2 \right) \frac{d}{d\mu} \right] - \frac{s^2}{1 - \mu^2} - \frac{s}{\varpi} + \epsilon \left(\varpi^2 - \mu^2 \right) - \frac{2 \epsilon \mu \varpi \left(\varpi\hat{D} - s\mu \right)}{s^2 - \epsilon \varpi^2 \left( 1 - \mu^2 \right)},
\label{eq:fundamental_eqn_operator}
\end{equation}
in agreement with equation (7.8) of LH68 (see Appendix \ref{append:identity}).

\subsection{B.2. Near-Equator Solutions ($\epsilon \gg 1$)}
\label{subsect:classical_spherical_equator}

When $\epsilon \gg 1$, the last term in equation (\ref{eq:fundamental_eqn_operator}) is small compared to the other terms.  The first and second terms cannot be dropped because the boundary conditions at $\mu = \pm 1$ have to be satisfied (LH68)\footnote{Specifically, the spheroidal wave functions need to be finite at the poles (Abramowitz \& Stegun 1970).}.  In this approximation, one may recognize $\hat{L}_s \tilde{v}_{\theta_0} = 0$ as the ``spheroidal wave equation'' with the solutions ${\cal S}_{sn}\left(\sqrt{\epsilon},\mu\right)$ (Abramowitz \& Stegun 1970), where the ``separation constant'' is
\begin{equation}
\Lambda_{sn}\left(q\right) = \epsilon \varpi^2 - \frac{s}{\varpi}
\label{eq:separation}
\end{equation}
and $q=\sqrt{\epsilon}$.  Note that for $l \equiv n - s$,
\begin{equation}
\frac{\Lambda_{sn}\left(q\right)}{q} \rightarrow 2l + 1
\end{equation}
for $q \gtrsim 4$ (Abramowitz \& Stegun 1970).  When $\epsilon=0$, $\hat{L}_s$ becomes the operator for the associated Legendre equation; the corresponding solution is ${\cal P}^{(s)}_n (\mu)$ and the separation constant is $\Lambda_{sn} = n(n+1)$ (Arfken \& Weber 1995).  In this case, only the eastward-propagating Poincar\'{e} mode exists.  When $\epsilon \ne 0$, a westward-propagating Poincar\'{e} mode appears in addition to the Rossby mode.  The dispersion relation is extracted from equation (\ref{eq:separation}):
\begin{equation}
\varpi^3 - \varpi \Lambda_{sn} \left(\frac{\lambda_{\rm R}}{R}\right)^2 - s\left(\frac{\lambda_{\rm R}}{R}\right)^2 = 0.
\label{eq:dispersion_spherical}
\end{equation}

The Poincar\'{e} and Rossby waves are described by the so-called ``Type 1'' and ``Type 2'' solutions of LH68, respectively.  The azimuthal velocity perturbation in both cases has the functional form,
\begin{equation}
v^\prime_{\theta} \approx -i \tilde{\eta}_0 ~{\cal H}_l \Psi,
\end{equation}
where ${\cal H}_l(X)$ is the Hermite polynomial and
\begin{equation}
\begin{split}
&X \equiv \epsilon^{1/4} \mu,\\
&\Psi \equiv \exp{\left(-\frac{X^2}{2} \right)} ~\exp{i \left(s\phi - \omega t \right)}.\\
\end{split}
\end{equation}

We first focus on the Type 1 solutions, where 
\begin{equation}
\varpi \sim \pm \frac{\sqrt{2l+1}}{\epsilon^{1/4}}.
\end{equation}
Making use of the second and third equations in the set (\ref{eq:amplitudes_spherical}) leads to
\begin{equation}
\eta_0 = \frac{\left( \varpi \hat{D} - s\mu \right) \tilde{v}_{\theta_0}}{\varpi^2 \epsilon \left(1 - \mu^2 \right) - s^2} \approx \frac{\hat{D} \tilde{v}_{\theta_0}}{\varpi \epsilon}.
\end{equation}
The preceding approximation is justified because $s\mu \tilde{v}_{\theta_0} \ll \varpi\hat{D} \tilde{v}_{\theta_0}$, and $\varpi^2 \epsilon \propto \epsilon^{1/2} \gg s^2$.  As we will see later, these approximations are inviable for the Type 2 solutions.  It follows that
\begin{equation}
h^\prime \approx \pm \frac{\tilde{h}_0}{\sqrt{2l+1}} ~\epsilon^{-1/2} \left( l {\cal H}_{l-1} - \frac{1}{2} {\cal H}_{l+1} \right) \Psi.
\end{equation}
Notice that we have the ``$\pm$'' instead of the ``$\mp$'' sign in equation (8.32) of LH68, because we are defining $\theta$ as the co-latitude.  

To obtain $v^\prime_\phi$, we use the second equation in (\ref{eq:amplitudes_spherical}),
\begin{equation}
v_{\phi_0} \approx -\frac{\mu \tilde{v}_{\theta_0}}{\varpi}.
\end{equation}
The preceding approximation is valid because $\mu \tilde{v}_{\theta_0} \propto \epsilon^{-1/4} \gg s\eta \propto \epsilon^{-1/2}$.  It follows that
\begin{equation}
v^\prime_{\phi} \approx \mp \frac{\tilde{\eta}_0}{\sqrt{2l + 1}} \left(l {\cal H}_{l-1} + \frac{1}{2} {\cal H}_{l+1} \right) ~\Psi.
\end{equation}

Obtaining the Type 2 solutions, where
\begin{equation}
\varpi \sim -\frac{s}{\left(2l+1 \right) \sqrt{\epsilon}},
\end{equation}
involves realizing that the approximations made to obtain the Type 1 solutions do not hold, because
\begin{equation}
\begin{split}
&s\mu \tilde{v}_{\theta_0} \propto \epsilon^{-1/4} \sim  \varpi\hat{D} \tilde{v}_{\theta_0} \propto \epsilon^{-1/4},\\
&\varpi^2 \epsilon \sim s^2 ,\\
&\frac{\mu \tilde{v}_{\theta_0}}{\varpi} \propto \epsilon^{1/4} \sim \frac{s \eta_0}{\varpi} \propto \epsilon^{1/4},\\
\end{split}
\end{equation}
and therefore the previously dropped terms need to be retained.  Making use of the recursion relations for Hermite polynomials, we arrive at:
\begin{equation}
\begin{split}
&h^\prime \approx \tilde{h}_0 ~\frac{2l+1}{2s} ~\epsilon^{-1/4} ~\left( {\cal H}_{l-1} +  \frac{1}{2l+2}{\cal H}_{l+1} \right) ~\Psi,\\
&v^\prime_{\phi} \approx - \tilde{\eta}_0 ~\frac{2l+1}{2s} ~\epsilon^{1/4} ~\left( {\cal H}_{l-1} -  \frac{1}{2l+2}{\cal H}_{l+1} \right) ~\Psi.\\
\end{split}
\end{equation}
Again, notice that the plus and minus signs are switched compared to LH68.  Note that we require $l \ge 0$ and $l \ge 1$ for Type 1 and 2 solutions, respectively.

\subsection{B.3. Near-Pole Solutions ($-\epsilon \gg 1$)}
\label{subsect:classical_spherical_pole}

LH68 showed that $\varpi \rightarrow \pm 1$ as $-\epsilon$ increases.  He reasoned that for the penultimate term in equation (\ref{eq:fundamental_eqn_operator}) to not dominate the other terms, we must have $- \epsilon \left( \varpi^2 - \mu^2 \right) \sim 1$.  When $-\epsilon \gg 1$,  $( 1 - \mu^2) \ll 1$ and $\mu^2 \sim 1$.  Therefore, the solutions to $\hat{L}_s \tilde{v}_{\theta_0} = 0$ are concentrated near the poles for large values of $-\epsilon$.

To proceed, one assumes $\varpi$ to have the form,
\begin{equation}
\varpi = \pm 1 + \frac{Q}{\sqrt{-\epsilon}}.
\end{equation}
Using the substitution,
\begin{equation}
Y \equiv \sqrt{-\epsilon} \left( 1 - \mu^2 \right),
\end{equation}
the equation $\hat{L}_s \tilde{v}_{\theta_0} = 0$ becomes
\begin{equation}
\left[ \frac{d^2}{dY^2} - \frac{1}{4} + \frac{Q}{2Y} - \frac{s^2 \pm 2s}{4Y^2} \right] \tilde{v}_{\theta_0} = 0,
\end{equation}
readily recognized as Whittaker's equation (Abramowitz \& Stegun 1970).  The solutions are
\begin{equation}
\tilde{v}_{\theta_0} \approx V_0 ~\exp{\left(-\frac{Y}{2}\right)} ~Y^{\left(m + 1 \right)/2} ~{\cal L}^{\left( m \right)}_l \left(Y\right),
\end{equation}
where ${\cal L}^{\left(m\right)}_l(Y)$ is the associated Laguerre polynomial (Arfken \& Weber 1995).  For $\varpi \approx \pm 1$, we have $m = \vert s \pm 1 \vert$; also, $Q = 2m + 2l + 1$.  The $\varpi \approx -1$ and $+1$ cases correspond to the ``Type 4'' and ``Type 5'' solutions of LH68, respectively.  The quantities $h^\prime$ and $v^\prime_{\phi}$ can be calculated using the procedure described in Appendix \ref{subsect:classical_spherical_equator} to compute the Type 2 solutions.

\section{C. Generalized Identity of Longuet-Higgins (1968)}
\label{append:identity}

We provide a proof for the identity used to go from equation (7.5) to (7.8) of LH68 and generalize it for MHD systems.  Using the test functions ${\cal F} = {\cal F}(\mu)$ and ${\cal G} = {\cal G}(\mu)$, one show can that
\begin{equation}
\left(\chi \hat{D} + \mu s\right)\left[{\cal G} \left(\chi \hat{D} - \mu s\right) \right]{\cal F} = {\cal G} \left(\chi \hat{D} + \mu s\right) \left(\chi \hat{D} - \mu s\right){\cal F} + \chi \left(\hat{D}{\cal G}\right) \left[\chi \left(\hat{D}{\cal F}\right) - \mu s {\cal F} \right],
\end{equation}
since
\begin{equation}
\left(\chi \hat{D} + \mu s\right) \left(\chi \hat{D} - \mu s\right){\cal F} = \chi^2 \hat{D} \left(\hat{D}{\cal F} \right) - \chi \hat{D}\left(\mu s {\cal F} \right) + \mu s \chi \left(\hat{D}{\cal F} \right) - \left(\mu s \right)^2 {\cal F}.
\end{equation}
Now, we consider
\begin{equation}
{\cal G} = \left[ \varpi \chi \epsilon \left( 1 - \mu^2 \right) - s^2 \right]^{-1},
\end{equation}
such that
\begin{equation}
\left(\chi \nabla^2_{\rm LH} - s \right){\cal F} + \frac{2 \varpi \chi \epsilon \mu}{ \varpi \chi \epsilon \left( 1 - \mu^2 \right) - s^2} \left(\chi \hat{D} - \mu s\right){\cal F} + \varpi \epsilon \left(\chi^2 - \mu^2 \right){\cal F} = 0.
\end{equation}
The operator used in Longuet-Higgins (1965) and LH68, originally denoted by $\nabla^2$, is defined by us as
\begin{equation}
\nabla^2_{\rm LH} \equiv \frac{d}{d \mu}\left[\left( 1 - \mu^2 \right) \frac{d}{d\mu} \right] - \frac{s^2}{1 - \mu^2}.
\end{equation}
To use these identities for non-magnetized systems, simply set $\chi = \varpi$.

\section{D. $\beta$-Plane Treatment for MHD Equatorial Eigenfunctions}
\label{append:betaplane}

The equatorial eigenfunctions can be obtained in the $\beta$-plane approximation, an analysis first performed by Matsuno (1966), who recognized it as an eigenvalue problem.  The $\beta$-plane analysis is algebraically more tractable and provides a consistency check on the results in \S\ref{subsect:mhd_spherical_equator}.  We first normalize the equations using the length and time scales:
\begin{equation}
L = \sqrt{\frac{c_0}{\beta}}, ~T = \frac{1}{\sqrt{c_0 \beta}}.
\end{equation}
We also let $\tilde{\gamma}_0 \equiv g\tilde{h}_0/c^{3/2} \beta^{1/2}$, where $\tilde{h}_0 \equiv h_0/L$.  Subsequently, quantities in this section marked with a ``tilde'' are normalized by $L$ and/or $T$, unless otherwise specified.

Seeking $\exp{i(k_x x - \omega t)}$ wave solutions, the amplitude equations become:
\begin{equation}
\begin{split}
&\tilde{y} \tilde{v}_{x_0} - i \tilde{\chi} \tilde{v}_{y_0} + \frac{d \tilde{\gamma}_0}{d \tilde{y}} = 0,\\
&-i \tilde{\chi} \tilde{v}_{x_0} - \tilde{y} \tilde{v}_{y_0} + i \tilde{k}_x \tilde{\gamma}_0  = 0,\\
&i \tilde{k}_x \tilde{v}_{x_0} + \frac{d \tilde{v}_{y_0}}{d \tilde{y}} - i \tilde{\omega} \tilde{\gamma}_0  = 0,\\
\end{split}
\label{eq:mhdbeta_amp}
\end{equation}
where
\begin{equation}
\begin{split}
& \tilde{\chi} \equiv \tilde{\omega} - \frac{\tilde{B}^2_z}{\tilde{\omega}},\\
&\tilde{B}_z \equiv \frac{B_z}{\sqrt{4 \pi \rho \beta c_0} H}.\\
\end{split}
\end{equation}

Manipulating the second and third equations in (\ref{eq:mhdbeta_amp}) yields:
\begin{equation}
\begin{split}
&\tilde{v}_{x_0} = \frac{1}{i\left(\tilde{\omega}\tilde{\chi} - \tilde{k}^2_x \right)} \left( \tilde{k}_x \frac{d \tilde{v}_{y_0}}{d \tilde{y}} - \tilde{\omega} \tilde{y} \tilde{v}_{y_0} \right),\\
&\tilde{\gamma}_0 = - \frac{1}{i\left(\tilde{\omega}\tilde{\chi} - \tilde{k}^2_x \right)} \left( - \tilde{\chi} \frac{d \tilde{v}_{y_0}}{d \tilde{y}} + \tilde{k}_x \tilde{y} \tilde{v}_{y_0} \right).\\
\end{split}
\end{equation}
Substituting the preceding expressions into the first equation of (\ref{eq:mhdbeta_amp}) gives:
\begin{equation}
\left[ \frac{d^2}{d \tilde{Y}^2} - \alpha^{1/2} \tilde{y}^2 + \Lambda_{sn} \alpha^{-1/2} \right] \tilde{v}_{y_0} = 0.
\end{equation}
As expected, the governing equation for $\tilde{v}_{y_0}$ is the parabolic cylinder equation, where $\alpha \equiv \tilde{\omega}/\tilde{\chi}$ and $\tilde{Y} \equiv \alpha^{1/4} \tilde{y}$.  It has the solution:
\begin{equation}
\begin{split}
&\tilde{v}_y = V_0 ~\tilde{{\cal H}}_l ~\tilde{\Psi},\\
&\tilde{\Psi} \equiv \exp{\left(- \frac{\tilde{Y}^2}{2} \right)} ~\exp{i(k_x x - \omega t)}.\\
\end{split}
\end{equation}
Here, $\tilde{{\cal H}}_l \equiv {\cal H}_l(\tilde{Y})$ and $V_0$ is an arbitrary normalization constant.

The dispersion relation is:
\begin{equation}
\begin{split}
&\tilde{\omega}^8 - 2\left(2 \tilde{B}^2_z + \tilde{k}^2_x \right) \tilde{\omega}^6 - 2\tilde{k}_x \tilde{\omega}^5\\
&+\left[ 6\tilde{B}^2_z \left(\tilde{B}^2_z + \tilde{k}^2_x \right) +  \tilde{k}^4_x - \left( 2l + 1 \right)^2 \right] \tilde{\omega}^4\\
&+ \left[ -2 \tilde{B}^2_z \left(2 \tilde{B}^4_z + \tilde{k}^4_x \right) + \tilde{k}^2_x \left(1 - 6\tilde{B}^4_z \right) + \tilde{B}^2_z \left(2l+1\right)^2 \right] \tilde{\omega}^2\\
&+ 2 \tilde{k}_x \left(2 \tilde{B}^2_z + \tilde{k}^2_x \right) \tilde{\omega}^3 -  2 \tilde{k}_x \tilde{B}^2_z \left(\tilde{B}^2_z + \tilde{k}^2_x \right) \tilde{\omega}\\
&+ \tilde{B}^4_z \left( \tilde{B}^4_z + 2 \tilde{B}^2_z \tilde{k}^2_x + \tilde{k}^4_x \right) = 0.\\
\end{split}
\end{equation}

For the magneto-Rossby and west magneto-Poincar\'{e} modes, the other eigenfunction solutions are:
\begin{equation}
\begin{split}
&\tilde{v}^\prime_x = \frac{V_0}{i\left(\tilde{\omega}\tilde{\chi} - \tilde{k}^2_x \right)} \left[ l\tilde{{\cal H}}_{l-1} {\cal B}_+ - \frac{1}{2} \tilde{{\cal H}}_{l+1} {\cal B}_- \right] ~\tilde{\Psi},\\
&\tilde{\gamma}^\prime = - \frac{V_0}{i\left(\tilde{\omega}\tilde{\chi} - \tilde{k}^2_x \right)} \left[ l\tilde{{\cal H}}_{l-1} {\cal A}_+ + \frac{1}{2} \tilde{{\cal H}}_{l+1} {\cal A}_- \right] ~\tilde{\Psi},\\
\end{split}
\end{equation}
where $\tilde{\gamma}^\prime \equiv g\tilde{h}^\prime/c^{3/2} \beta^{1/2}$.  For the east magneto-Poincar\'{e} mode, we have:
\begin{equation}
\begin{split}
&\tilde{v}^\prime_x = \frac{V_0}{i\left(\tilde{\omega}\tilde{\chi} - \tilde{k}^2_x \right)} \left[ l\tilde{{\cal H}}_{l-1} {\cal B}_- - \frac{1}{2} \tilde{{\cal H}}_{l+1} {\cal B}_+ \right] ~\tilde{\Psi},\\
&\tilde{\gamma}^\prime = - \frac{V_0}{i\left(\tilde{\omega}\tilde{\chi} - \tilde{k}^2_x \right)} \left[ l\tilde{{\cal H}}_{l-1} {\cal A}_- + \frac{1}{2} \tilde{{\cal H}}_{l+1} {\cal A}_+ \right] ~\tilde{\Psi}.\\
\end{split}
\end{equation}
Analogous to the case of spherical geometry, we have:
\begin{equation}
\begin{split}
&{\cal A}_\pm \equiv \tilde{k}_x \left( \frac{\tilde{\chi}}{\tilde{\omega}} \right)^{1/4} \pm \left\vert \tilde{\chi} \right\vert^{3/4} \left\vert \tilde{\omega} \right\vert^{1/4},\\
&{\cal B}_\pm \equiv \tilde{k}_x \left( \frac{\tilde{\omega}}{\tilde{\chi}} \right)^{1/4} \pm \left\vert \tilde{\chi} \right\vert^{1/4} \left\vert \tilde{\omega} \right\vert^{3/4}.\\
\end{split}
\end{equation}
The (normalized) magnetic field perturbations are obtained from $\tilde{b}_{x,y} = i \tilde{B}_z \tilde{v}^\prime_{x,y}/ \omega$.




\begin{references}

\reference{} Abramowitz, M., \& Stegun, I.A. \ 1970, Handbook of Mathematical Functions, Ninth Printing (New York: Dover)

\reference{} Arfken, G.B., \& Weber, H.J. \ 1995, Mathematical Methods for Physicists, 4th Edition (San Diego: Academic Press)

\reference{} Berkhout, R.G., \& Levin, Y. \ 2008, MNRAS, 385, 1029

\reference{} Braginsky, S.I. \ 1998, Earth Planets Space, 50, 641

\reference{} Cumming, A., \& Bildsten, L. \ 2000, ApJ, 544, 453

\reference{} Draine, B.T. \ 1986, MNRAS, 220, 133

\reference{} Gill, A.E. \ 1982, Atmosphere-Ocean Dynamics (Academic Press, Inc.) [G82]

\reference{} Gilman, P.A. \ 2000, ApJ, 544, L79

\reference{} Heyl, J.S. \ 2004, ApJ, 600, 939

\reference{} Holton, J.R. \ 2004, An Introduction to Dynamic Meteorology, 4th Edition (Elsevier Academic Press)

\reference{} Inogamov, N.A., \& Sunyaev, R.A. \ 1999, Astron. Lett., 25, 269

\reference{} Ivanov, M. I.. \ 2007, Fluid Dynamics, 42, 644

\reference{} Kundu, P.K., \& Cohen, I.M. \ 2004, Fluid Mechanics, 3rd Edition (Elsevier Academic Press) [KC04]

\reference{} Longuet-Higgins, M.S. \ 1965, Proc. Roy. Soc. A, 284, 40

\reference{} Longuet-Higgins, M.S. \ 1968, Phil. Trans. Roy. Soc., 262, 511 [LH68]

\reference{} Margules, M. \ 1893, Sber. Akad. Wiss. Wien, 102, 11

\reference{} Matsuno, T. \ 1966, J. Meteorological Soc. of Japan, 44, 25

\reference{} Pedlosky, J. \ 1987, Geophysical Fluid Dynamics (New York: Springer)

\reference{} Schecter, D.A., Boyd, J.F., \& Gilman, P.A. \ 2001, ApJ, 551, L185

\reference{} Spitkovsky, A., Levin, Y., \& Ushomirsky, G. \ 2002, ApJ, 566, 1018

\reference{} Strohmayer, T.E., Zhang, W., \& Swank, J.H. \ 1998, ApJ, 503, L147

\reference{} Strohmayer, T.E., \& Bildsten, L. \ 2006, in Compact Stellar X-Ray Sources, ed. W.H.G. Lewin and M. van der Klis (Cambridge University Press), 113 (arXiv:astro-ph/0301544v2)

\reference{} van der Klis, M. \ 2000, ARA\&A, 2000, 38, 717

\reference{} Zaqarashvili, T.V., Oliver, R., Ballester, J.L., \& Shergelashvili, B.M. \ 2007, A\&A, 470, 815

\reference{} Zaqarashvili, T.V., Oliver, R., \& Ballester, J.L. \ 2009, ApJ, 691, L41

\end{references}
\end{document}